\documentclass[aps,prb,reprint,twocolumn,amsmath,amssymb,nofootinbib]{revtex4-1}

\usepackage{graphicx}
\usepackage{dcolumn}
\usepackage{bm}
\usepackage{braket}

\DeclareMathOperator{\R}{Re}
\DeclareMathOperator{\I}{Im}
\DeclareMathOperator{\Tr}{Tr}
\DeclareMathOperator{\adj}{adj}

\begin{document}

\preprint{APS/123-QED}

\title{Unified theory of resonances and bound states in the continuum in 
Hermitian tight-binding models}

\author{A.\,A.\,Gorbatsevich}
 \affiliation{P.N. Lebedev Physical Institute of the Russian Academy of Sciences, 119991, Moscow, Russia.}
 \altaffiliation[Also at ]{National Research University of Electronic Technology, 124498, Zelenograd, Moscow, Russia.}
 \email{aagor137@mail.ru}

\author{N.\,M.\,Shubin}
\affiliation{National Research University of Electronic Technology, 124498, Zelenograd, Moscow, Russia.}

\date{\today}

\begin{abstract}

We study transport properties of an arbitrary two terminal Hermitian system within a tight-binding approximation and derive the expression for the transparency in the form, which enables one to determine exact energies of perfect (unity) transmittance, zero transmittance (Fano resonance) and bound state in the continuum (BIC). These energies correspond to the real roots of two energy-dependent functions that are obtained from two non-Hermitian Hamiltonians: the Feshbach's effective Hamiltonian and the auxiliary Hamiltonian, which can be easily deduced from the effective one. BICs and scattering states are deeply connected to each other. We show that transformation of a scattering state into a BIC can be formally described as a ``phase transition'' with divergent generalized response function. Design rules for quantum conductors and waveguides are presented, which determine structures exhibiting coalescence of both resonances and antiresonances resulting in the formation of almost rectangular transparency and reflection windows. The results can find applications in construction of molecular conductors, broad band filters, quantum heat engines and waveguides with controllable BIC formation.

\end{abstract}

\pacs{Valid PACS appear here}
\maketitle

\section{Introduction}
Resonances play the central role in the physics of open quantum systems and waveguides. \cite{bib:MoiseyevBook, bib:Monti2017,bib:FanoRev} Therefore, the ability to design structures with required resonance properties is of the primary importance for the whole field of nanoelectronic and nanophotonic engineering. Past years demonstrated a steady progress in understanding properties of open quantum systems and subwavelength electronic and optical structures.\cite{bib:MoiseyevBook,bib:Hsu,bib:Monti2017,bib:FanoRev} In single connected structures the main type of resonances are  Fabry-Perot (FP) or Breit-Wigner (BW) resonance. \cite{bib:LandauQM} In multiple connected quantum systems interplay of different scattering paths can result in both constructive (resonances) and destructive (antiresonances) interference. A typical example of such structure is an Aharonov-Bohm interferometer. If one of the paths includes a (quasi) localized state, then an asymmetric Fano-Feshbach resonance is formed.\cite{bib:FanoRev} In molecular or quantum dot (QD) multiple connected quantum conductors (e.g. QD rings) all the paths consist of (quasi) localized states and antiresonances can be considered as resonances of Fano-Feshbach type.\cite{bib:QDR1,*bib:QDR2,*bib:QDR3} At the antiresonance the transparency can turn exactly into zero and, hence, in the region between two scatterers exhibiting antiresonances FP resonator is formed and a wave is trapped. Such a state is, in fact, the bound state in the continuum (BIC).\cite{bib:Hsu} Existence of BICs was proposed on the eve of the quantum mechanics\cite{bib:BIC1929} but only recently BIC has been recognized as a wave phenomena\cite{bib:Wintgen1985} and a variety of approaches to realize BIC was proposed\cite{bib:Hsu} and experimentally verified\cite{bib:Hsu,bib:BICLaser} with BIC in FP resonator being the simplest one.

A common way to describe resonance characteristics of a system is the scattering matrix language.\cite{bib:Fesh1,*bib:Fesh2,*bib:Fesh3,bib:LandauQM} Scattering matrix amplitude is often expressed in terms of the effective non-Hermitian Hamiltonian, which can be obtained via Feshbach's projection technique.\cite{bib:Fesh1,*bib:Fesh2,*bib:Fesh3} Resonances correspond to the poles of the scattering matrix in the lower half of the complex energy plane\cite{bib:Fesh1,bib:LandauQM,bib:Sasada2011} or equivalently -- to the eigenvalues of the Feshbach's effective Hamiltonian.  For a narrow and isolated resonance its location almost perfectly coincides with the real part of the scattering matrix ($S$-matrix) pole. For wide and/or interacting (closely spaced) resonances this is not true. \cite{bib:Gor,bib:DoublePole2,bib:Belozerova} Interaction of resonances can result in their coalescence (collapse of resonances\cite{bib:Gor,bib:Pastawski2008}) that was described for semiconductor heterostructures with two resonances\cite{bib:Romo,bib:Gor} and for tunneling quantum dots with three resonances.\cite{bib:rotter1995,bib:GorShAop} Collapse of eigenmodes was also observed in semiconductor cavities with Rabi splitting,\cite{bib:Collapse1} classical electrical circuits,\cite{bib:EPelect,bib:EPelect2} in quantum tunneling structures\cite{bib:EPQD1,bib:EPQD2} etc. However coalescence of resonances can not be described in the terms of the scattering matrix poles behavior. \cite{bib:Gor,bib:GorShAop}

Implementation of novel physical concepts such as $\mathcal{PT}$-symmetry and $\mathcal{PT}$-symmetry breaking,\cite{bib:PTBend98,*bib:PTBend99,*bib:PTBend07} (here $\mathcal{P}$ stands for the space inversion and $\mathcal{T}$ -- for the time reversal symmetries) opened new directions in study of open electronic systems with non-Hermitian Hamiltonians\cite{bib:MoiseyevBook,bib:PTBend07} and electromagnetic waveguide structures with combined gain and loss.\cite{bib:PTopt1,bib:PTOptNat} It also shed a light on the mechanism of the coalescence of resonances. In Refs.\cite{bib:AnnPTScat,bib:PTScat,bib:Jin2010,bib:Jin2011,bib:GorShJETP,bib:GorShAop} the relation of resonances in quantum conductors to the $\mathcal{PT}$-symmetry was studied. In fact, under the condition of the perfect resonance incoming and outgoing particle flows are related to each other by the $\mathcal{PT}$-symmetry. Therefore, the $\mathcal{PT}$-symmetry is inherent to the perfect resonance condition. 

An important feature of $\mathcal{PT}$-symmetric systems is the $\mathcal{PT}$-symmetry breaking phenomenon, which takes place at a some point in the parameter space, where two real eigenvalues coalesce and with further parameter variation turn into a pair of complex conjugated eigenvalues with nonzero imaginary parts.\cite{bib:PTBend98} Such a point in the parameter space is known as an exceptional point (EP). \cite{bib:BookKato,bib:Heiss2004,*bib:Heiss2012,bib:Berry,bib:Rotter2009,*bib:Rotter2015} The Hamiltonian at the EP takes the form of a Jordan matrix (which is obviously non-Hermitina). Eigenvectors coalesce at the EP into a single nondegenerate state\cite{bib:BookKato,bib:Rotter2009,*bib:Rotter2015,bib:EPWier} contrary to the crossing (diabolic) point, where they are degenerate and can be made orthogonal. It should be noted that, in general, the $\mathcal{PT}$-symmetry of the non-Hermitian Hamiltonian is not a necessary and sufficient condition of energy spectrum to be real.\cite{bib:Mostafa1,bib:KivshNJP} Because of the close mathematical similarity of Schroedinger and wave equations $\mathcal{PT}$-symmetry can be straightforwardly realized in optics, where $\mathcal{PT}$-symmetric terms correspond to the gain and loss regions. $\mathcal{PT}$-symmetry breaking and EPs have been demonstrated in coupled waveguides,\cite{bib:PTopt1} photonic lattices,\cite{bib:NatPTLight} $\mathcal{PT}$-symmetric plasmonic metamaterials,\cite{bib:Alaeian,bib:AnnMeta} lasers,\cite{bib:Laser,*bib:nature} coherent perfect absorbers\cite{bib:LonghiCPA} and other optical systems.

In fermionic system time-odd terms in the Hamiltonian destroys unitarity. Hence, the only way to realize non-Hermitian terms is to consider in-flow and out-flow processes in a dissipationless open quantum system as it has been done in Ref.\cite{bib:Jin2010}. The same authors showed\cite{bib:Jin2011} that scattering state of an arbitrary Hermitian lattice can be described as an eigenstate of an auxiliary non-Hermitian Hamiltonian with imaginary terms that describe incoming and outgoing particle flows. Recently within the framework of tight-binding approximation we obtained the exact expression for the transparency of a dissipationless quantum chain (single connected quantum conductor),\cite{bib:GorShJETP,bib:GorShAop} which directly relates the transparency maxima to the eigenvalues of an auxiliary non-Hermitian Hamiltonian that can be straightforwardly deduced from the Feshbach's effective Hamiltonian. In spatially symmetric systems auxiliary Hamiltonian becomes $\mathcal{PT}$-symmetric and possesses real eigenvalues, which exactly determine the location of perfect resonances. At the EP of the auxiliary Hamiltonian resonances coalesce and the broad transparency window is formed. Transparency at EP has essentially a non-BW profile.\cite{bib:Gor,bib:GorShJETP} Poles of the scattering matrix (Green's function) can also coalesce resulting in the formation of a double pole.\cite{bib:Heiss2014,bib:DoublePole1,bib:DoublePole2} However, its location, in general, has no direct relation to the coalescence of resonances and, hence, to physical observables. \cite{bib:Gor,bib:GorShAop} Although, as was shown in Refs.\cite{bib:Ambi,bib:Cho}, physical properties of the system do change at EP of the scattering matrix of a system with balanced gain and loss, where two unimodular eigenvalues of the $S$-matrix turn into two non-unimodular. In Ref.\cite{bib:Heiss2014} interaction of Fano resonances was analyzed in connection with the formation of the Green's function double poles for interacting scattering channels, but just as in the case of BW resonances location of a double pole has, in general, no relation to the location of the coalescence of resonances.

Energy (frequency) of a BIC is the real eigenvalue of the effective Hamiltonian.\cite{bib:Hsu,bib:BulgJETP} Because of unitarity of the $S$-matrix (in nodissipative system) some relations should exist between zeroes of the denominator (eigenvalues of the effective Hamiltonian) and zeroes of the numerator (antiresonances) in the expression for the $S$-matrix, which has been  recently studied on phenomenological grounds in Ref.\cite{bib:BICDiverg}. On the other hand, zeroes of the effective Hamiltonian in complex plane determine resonances. Hence, resonances and BIC energies (frequencies) are related to each other as well. In this paper we present microscopic theory, which provides a unified description of transparency maxima (resonances), transparency zeroes (antiresonances) and BIC energies (frequencies).

Structure of the paper is as follows. The model under consideration is described in Sec.~\ref{Sec.Sys}. In Sec.~\ref{Sec.Trans} within a tight-binding approximation we derive a generalized formula for the transparency of an arbitrary two terminal multiple connected molecular (or QD) conductor or waveguide. This formula depends only on two functions of energy,  zeroes of which determine resonances and antiresonances.  In Sec.~\ref{Sec.BIC} we study situation when zeros of both functions coincide and BIC is formed. We show that the transition to a BIC state in the parameter space is characterized by the singularity in generalized response function just as in the case of the second order phase transition. However, the formation of the BIC state is discontinuous. In Sec.~\ref{Sec.ColZer} we present models that demonstrate the coalescence of Fano resonances either at EP or crossing points accompanied by the formation of wide reflection windows. Summary is made in Sec.~\ref{Sec.Sum}.

\section{General relations}
\label{Sec.Sys}
\subsection{Model}
We consider an arbitrary $N$-site Hermitian structure connected to two semi-infinite leads. Every isolated site is assumed to have a single localized state with a real energy $\varepsilon_{i}$. Within a tight-binding approximation this system can be described by the Hamiltonian
\begin{equation}
\hat H=\hat H_{0}+\hat H_{L}+\hat H_{R}+\hat H_{int}^{L}+\hat H_{int}^{R}.
\label{eq:HTot}
\end{equation}
The first term in~(\ref{eq:HTot}) describes isolated Hermitian $N$-site structure:
\begin{equation}
\hat H_{0}=\sum_{i=1}^{N}{\varepsilon_{i}a_{i}^{\dag}a_{i}}+\sum_{i,j=1,i<j}^{N}{\left(\tau_{ij}a_{j}^{\dag}a_{i}+h.c.\right)},
\label{eq:H0}
\end{equation}
where $a_{i}^{\dag}(a_{i})$ is the creation (annihilation) operator of the electron on the $i$-th site and $\tau_{ij}$ is the tunneling matrix element between the $i$-th and the $j$-th sites. The problem we consider here is single electron, without taking into account many-body effects such as electron-electron interaction, electron-phonon scattering and etc. Neglecting electron-electron Coulomb and exchange interaction limits us to the case of small on-site amplitudes and fast tunneling rates in order to prevent charge accumulation. 

The Hamiltonian~(\ref{eq:HTot}) is also applicable to the description of optical waveguide systems within an evanescent wave coupling approximation. In an optical system we consider a light propagation along waveguides (instead of a time evolution in a quantum system), on-site energies and tunneling matrix elements are replaced by corresponding propagation constants and evanescent field overlapping integrals.\cite{bib:okamoto, bib:LonghiRev} In contrast to electron systems, here there is no restriction on the field amplitudes.

Contacts with the spectrum $\varepsilon_{L(R)}=\varepsilon_{L(R)}(p)$ are described by the Hamiltonians $\hat H_{L}$ and $\hat H_{R}$:
\begin{equation}
\hat H_{L(R)}=\sum_{p}{\varepsilon_{L(R)}(p)a_{L(R),p}^{\dag}a_{L(R),p}}.
\label{eq:HCont}
\end{equation}
Operator $a_{L(R),p}$ in~(\ref{eq:HCont}) corresponds to the state in the left (right) contact with momentum $p$. Term $\hat H_{int}^{L(R)}$ in Eq.~(\ref{eq:HTot}) describes interaction between the state with momentum $p$ in the left (right) contact and the $i$-th site of the structure: 
\begin{equation}
\hat H_{int}^{L(R)}=\sum_{p,i}{\left(\gamma^{L(R)}_{p,i}a_{i}^{\dag}a_{L(R),p}+h.c.\right)}.
\label{eq:HSysContInt}
\end{equation}
Here $\gamma^{L(R)}_{p,i}$ is a matrix element, which, in general, is energy and momentum dependent.
 
\subsection{Transmission coefficient}
Transport properties of the quantum system are defined by its transmission coefficient, which can be written as:\cite{bib:Caroli1971,bib:DattaBook1997}
\begin{equation}
T=4\Tr{\left(\hat\Gamma^{R}\hat G^{r}\hat\Gamma^{L}\hat G^{a}\right)}.
\label{eq:TransDef}
\end{equation}
Here $\hat G^{r}$ and $\hat G^{a}=(\hat G^{r})^{\dag}$ are correspondingly retarded and advanced Green's functions of the system with interaction with contacts being taken into account:
\begin{equation}
\hat G^{r}=\left(\omega\hat I-\hat H_{eff}\right)^{-1},
\label{eq:GFHeff}
\end{equation}
where $\hat I$ is the $N\times N$ identical matrix and $\hat H_{eff}$ is the effective Hamiltonian\cite{bib:Fesh1,*bib:Fesh2,*bib:Fesh3} of the structure:
\begin{equation}
\hat H_{eff}=\hat H_{0}+\hat\Sigma^{L}+\hat\Sigma^{R}.
\label{eq:Heff}
\end{equation}
Here $\hat\Sigma_{L(R)}$ is the self-energy of the left (right) contact. Hermitian matrix $\hat\Gamma^{L(R)}$ from Eq.~(\ref{eq:TransDef}) describes an anti-Hermitian part of the corresponding contact self-energy:
\begin{equation}
\hat\Sigma^{L(R)}=\hat\delta^{L(R)}-i\hat\Gamma^{L(R)}.
\label{eq:SigmaDOS}
\end{equation}

In the system we consider contact self-energy can be derived as\cite{bib:Caroli1971}
\begin{equation}
\Sigma^{L(R)}_{ij}=\sum_{p,p'}{\gamma^{L(R)}_{p,i}\left(\hat G^{r}_{L(R)}\right)_{pp'}}\gamma^{L(R)*}_{p,j},
\label{eq:Sigma}
\end{equation}
where $\hat G^{r}_{L(R)}$ is the retarded Green's function of the isolated left (right) contact, which is diagonal in the momentum representation:
\begin{multline}
\left(\hat G^{r}_{L(R)}\right)_{pp'}=\left[\left(\omega-\hat H_{L(R)}\right)^{-1}\right]_{pp'}\\
=\left(\omega-\varepsilon_{L(R)}(p)\right)^{-1}\delta_{pp'}.
\label{eq:ContGreen}
\end{multline}
Therefore, assuming that matrix elements $\gamma^{L(R)}_{p,i}=\gamma^{L(R)}_{i}(\varepsilon_{L(R)})$ depend on energy $\varepsilon_{L(R)}=\varepsilon_{L(R)}(p)$ rather than on  momentum $p$, Hermitian and anti-Hermitian parts of decomposition~(\ref{eq:SigmaDOS}) can be written as follows:
\begin{equation}
\begin{split}
\delta^{L(R)}_{ij}(\omega)&=p.v.\int{\frac{\gamma^{L(R)}_{i}(\omega')\gamma^{L(R)*}_{j}(\omega')\rho_{L(R)}(\omega')}{\omega-\omega'}d\omega'},\\
\Gamma^{L(R)}_{ij}(\omega)&=\pi \gamma^{L(R)}_{i}(\omega)\gamma^{L(R)*}_{j}(\omega)\rho_{L(R)}(\omega).
\label{eq:Delta_Gamma}
\end{split}
\end{equation}
Here $\rho_{L(R)}$ is the density of states in the left (right) contact.

Thus, the transmission coefficient of the structure becomes
\begin{equation}
T=\frac{4\sum_{i,j,m,k=1}^{N}{(-1)^{i+j+m+k}M_{ij}^{*}M_{mk}\Gamma^{R}_{jk}\Gamma^{L}_{mi}}}{\left|\det{\left(\omega\hat I-\hat H_{eff}\right)}\right|^{2}},
\label{eq:TransHeff}
\end{equation}
Minors $M_{ij}$ in Eq.~(\ref{eq:TransHeff}) are minors of the $(\omega\hat I-\hat H_{eff})$ matrix. In  Ref.\cite{bib:GorShAop} it was shown that in single connected quantum conductor the denominator and the numerator in Eq.~(\ref{eq:TransHeff}) are coupled to each other with simple relation, which makes it possible to determine exact positions of perfect transparency energies. Here we show that analogous decomposition of square module of the effective Hamiltonian characteristic determinant can be performed for an arbitrary multiple connected quantum conductor described by the model  (\ref{eq:HTot}). This property opens the way to separate control of transparency peaks and zeroes.

\section{Generalized formula for the transmission coefficient}
\label{Sec.Trans}
\subsection{Effective and auxiliary Hamiltonians and exact location of perfect and zero transmission energies}
According to Eq.~(\ref{eq:Delta_Gamma}) and in anl agreement with conventional approach for describing decay (see, e.g., Ref.~\cite{bib:HeffZelv}), matrix $\hat\Gamma^{L(R)}$ can be written as:
\begin{equation}
\hat\Gamma^{L(R)}=\mathbf{u}_{L(R)}\mathbf{u}_{L(R)}^{\dag},
\label{eq:Gam_a}
\end{equation}
with vector $(\mathbf{u}_{L(R)})_{i}=\sqrt{\pi\rho_{L(R)}}\gamma^{L(R)}_{i}$. Using Eq.~(\ref{eq:Gam_a}) we can simplify Eq.~(\ref{eq:TransDef}) in a way different from Eq.~(\ref{eq:TransHeff}). For brevity, let us introduce matrix
\begin{equation}
\hat A=\omega\hat I-\hat H_{0}-\hat\delta^{L}-\hat\delta^{R}.
\label{eq:A}
\end{equation}
Matrix $\hat A$ is Hermitian and this property is crucial in further calculations. Using $\hat A$ from Eq.~(\ref{eq:A}) we can simplify transmission coefficient to the following:
\begin{widetext}
\begin{multline}
T=4\Tr{\left\{\mathbf{u}_{R}\mathbf{u}_{R}^{\dag}\left(\hat A+i\mathbf{u}_{L}\mathbf{u}_{L}^{\dag}+i\mathbf{u}_{R}\mathbf{u}_{R}^{\dag}\right)^{-1}\mathbf{u}_{L}\mathbf{u}_{L}^{\dag}\left[\left(\hat A+i\mathbf{u}_{L}\mathbf{u}_{L}^{\dag}+i\mathbf{u}_{R}\mathbf{u}_{R}^{\dag}\right)^{-1}\right]^{\dag}\right\}}\\
=4\left|\mathbf{u}_{R}^{\dag}\left(\hat A+i\mathbf{u}_{L}\mathbf{u}_{L}^{\dag}+i\mathbf{u}_{R}\mathbf{u}_{R}^{\dag}\right)^{-1}\mathbf{u}_{L}\right|^{2}.
\label{eq:Simp1}
\end{multline}
\end{widetext}
Then applying Sherman–Morrison formula\cite{bib:SherMorr} and matrix determinant lemma\cite{bib:MatDetLem} we can simplify~(\ref{eq:Simp1}) and get:
\begin{equation}
T=\frac{4\left|\det{\hat A}\right|^{2}\left|\mathbf{u}_{R}^{\dag}\hat A^{-1}\mathbf{u}_{L}\right|^{2}}{\left|\det{\left(\hat A+i\mathbf{u}_{L}\mathbf{u}_{L}^{\dag}+i\mathbf{u}_{R}\mathbf{u}_{R}^{\dag}\right)}\right|^{2}}.
\label{eq:Simp2}
\end{equation}
According to Definitions~(\ref{eq:Gam_a}-\ref{eq:A}) the denominator of Eq.~(\ref{eq:Simp2}) is nothing more but the characteristic determinant of the effective Hamiltonian, which is also present in the denominator of Eq.~(\ref{eq:TransHeff}) for the transmission coefficient. Hence, the numerators of the equations (\ref{eq:Simp2}) and (\ref{eq:TransHeff}) should coincide with each other as well. From Eq.~(\ref{eq:Simp2}) it follows that the numerator of the transmission coefficient is a square module of some energy dependent quantity $P$, which is defined up to an arbitrary phase factor:
\begin{equation}
P=2\mathbf{u}_{R}^{\dag}\left(\adj{\hat A}\right)\mathbf{u}_{L}.
\label{eq:P}
\end{equation}
Here $\adj{\hat A}$ is an adjugate matrix of $\hat A$ from Eq.~(\ref{eq:A}).

Now we isolate term $4|\det{\hat A}|^{2}|\mathbf{u}_{R}^{\dag}\hat A^{-1}\mathbf{u}_{L}|^{2}=|P|^{2}$ in the denominator of Eq.~(\ref{eq:Simp2}) and then we just simplify the rest of the denominator. Applying matrix determinant lemma once again one can figure out that
\begin{multline}
\left|\det{\left(\omega\hat I-\hat H_{eff}\right)}\right|^{2}=\left|\det{\left(\hat A+i\mathbf{u}_{L}\mathbf{u}_{L}^{\dag}+i\mathbf{u}_{R}\mathbf{u}_{R}^{\dag}\right)}\right|^{2}\\
=\left|P\right|^{2}+\left|Q\right|^{2},
\label{eq:HeffPQ}
\end{multline}
where $Q$ is another function of $\omega$ defined up to an arbitrary phase factor:
\begin{equation}
Q=\det{\left(\hat A-i\hat\Gamma^{L}+i\hat\Gamma^{R}\right)}.
\label{eq:Q}
\end{equation}
Quantity $Q$ can be understood as a characteristic determinant of some auxiliary Hamiltonian $\hat H_{aux}$:
\begin{equation}
\hat H_{aux}=\hat H_{0}+\hat\delta^{L}+\hat\delta^{R}+i\hat\Gamma^{L}-i\hat\Gamma^{R}.
\label{eq:Haux}
\end{equation}
This auxiliary Hamiltonian differs from effective one~(\ref{eq:Heff}) only in the sign of $\hat\Gamma_{L}$ or $\hat\Gamma_{R}$. The choice of the sign is arbitrary, but for a sake of convenience, it can be assigned with an accordance to the direction of the current flow. Thus, the expression for the transmission coefficient of an arbitrary two-terminal Hermitian structure can be written in the following form:
\begin{equation}
T=\frac{\left|P\right|^{2}}{\left|P\right|^{2}+\left|Q\right|^{2}},
\label{eq:TransPQ}
\end{equation}
Equations~(\ref{eq:TransPQ}-\ref{eq:Haux}) represent the main result of this section. In fact we have proven the theorem that the transmission coefficient (\ref{eq:TransHeff},~\ref{eq:TransPQ}) can be expressed in terms of two characteristic determinants of two non-Hermitian Hamiltonians. One is the effective Hamiltonian (\ref{eq:Heff}), the other is the auxiliary Hamiltonian (\ref{eq:Haux}), which can be deduced from the effective one. Hence formula~(\ref{eq:TransPQ}) can be also written as:
\begin{equation}
T=\frac{\left|\omega\hat I-\hat H_{eff}\right|^{2}-\left|\omega\hat I-\hat H_{aux}\right|^{2}}{\left|\omega\hat I-\hat H_{eff}\right|^{2}}.
\label{eq:TransPQ2}
\end{equation}
Thus, according to the close relation between effective and auxiliary Hamiltonians [see Eqs.~(\ref{eq:Heff}) and~(\ref{eq:Haux})], one can see from Eq.~(\ref{eq:TransPQ2}) that transmission probability of the system can be fully described by the effective Hamiltonian only. It is worth mentioning that standard normalized Fano resonance profile \cite{bib:Fano1961}
\begin{equation}
T\left(\omega\right)=\frac{1}{1+q^{2}}\frac{\left(\omega+q\right)^{2}}{1+\omega^{2}}
\label{eq:FanoSTD}
\end{equation}
can be easily rewritten in the form~(\ref{eq:TransPQ}) with $|P|^{2}=(\omega+q)^{2}$ and $|Q|^{2}=(\omega q-1)^{2}$. 

The Hamiltonian $\hat H_{aux}$ depends on energy itself (because of self-energies) and, consequently, it's eigenvalue problem is non-linear and should be solved self-consistently. This fact can have a serious impact on its properties.\cite{bib:FanoEP} Strictly speaking, this means that $Q$ and $P$ are not polynomials, in general. However, if one neglects the energy dependence of self-energy, then $Q$ and $P$ can be cosidered as polynomials.\cite{bib:GorShAop}

In spatially symmetric structure non-Hermitian auxiliary Hamiltonian (\ref{eq:Haux}) becomes $\mathcal{PT}$-symmetric and possesses real eigenvalues.   
According to Eq.~(\ref{eq:TransPQ}), unity values of the transmission coefficient exactly coincide with real roots of $Q$, i.e. with real eigenvalues of the auxiliary Hamiltonian. Thus we have generalized the concept of the auxiliary Hamiltonian\cite{bib:GorShJETP,bib:GorShAop} on any arbitrary two-terminal structure. Moreover, Eq.~(\ref{eq:TransPQ}) enables one to determine exactly the positions of zero transmittance. Indeed, zero values of the transmission coefficient coincide with real roots of $P$. However, as it will be shown further, because roots of $P$ and $Q$ can coincide (which is just the case for a BIC) not all of the real roots of $Q$ correspond to the unity transmittance and not all real roots of $P$ correspond to zero transmittance.
At EP of auxiliary Hamiltonian its eigenvalues (roots of $Q$) coalesce, which results in the coalsecence of resonances. Below we show that in some cases roots of $P$ can be associated with eigenvalues of some Hermitian or $\mathcal{PT}$-symmetric Hamiltonian and can coalesce as well (either at crossing point or EP of the corresponding Hamiltonian) resulting in the coalescence of transparency zeroes.

It should be noted here that our approach has no restrictions both on the complex tunneling matrix elements $\tau_{ij}$ inside the structure and on the complex couplings $\gamma^{L(R)}_{i}$ with the contacts. Consequently, all phase shifts of hopping integrals $\Delta\phi=\frac{e}{\hbar}\int_{tunnel. path}{\mathbf{A}\cdot d\mathbf{l}}$ induced by external electromagnetic field with vector potential $\mathbf{A}$ can be taken into account properly allowing for the description of the Aharonov-Bohm effect.\cite{bib:ABLu} Thus, for instance, numerical analysis of some quantum dots based interferometers\cite{bib:ABQD1,bib:ABQD2,bib:ABQD3} can be extended to an explicit analytical description.

\subsection{Point contacts}
Expressions for the transmission coefficient can be simplified dramatically if we consider each lead interacting only with one site of the structure. Indeed, suppose that the left lead is attached to the site number $1$ and the right lead to the site number $N$. In this case each of the matrices $\hat\Gamma^{L(R)}$ and $\hat\delta^{L(R)}$ possess only one nonzero element each, namely:
\begin{equation}
\begin{split}
\Gamma^{L}_{11}&=\Gamma_{L}=\pi\left|\gamma^{L}_{1}(\omega)\right|^{2}\rho_{L}(\omega),\\
\Gamma^{R}_{NN}&=\Gamma_{R}=\pi\left|\gamma^{R}_{N}(\omega)\right|^{2}\rho_{R}(\omega),\\
\delta^{L}_{11}&=\delta_{L}=p.v.\int{\frac{\left|\gamma^{L}_{1}(\varepsilon)\right|^{2}\rho_{L}(\varepsilon)}{\omega-\varepsilon}d\varepsilon},\\
\delta^{R}_{NN}&=\delta_{R}=p.v.\int{\frac{\left|\gamma^{R}_{N}(\varepsilon)\right|^{2}\rho_{R}(\varepsilon)}{\omega-\varepsilon}d\varepsilon}.
\end{split}
\label{eq:Gam_Del}
\end{equation}
For the leads modeled by semi-infinite linear chains these quantities can be calculated explicitly.\cite{bib:Pastawski2008,bib:Sasada2011,bib:tightBindSelfE} Moreover, in this case the point contact approximation can be applied even if interaction with the lead is non-local, the only requirement is that only finite number of sites in the lead interact with the system. Suppose, that the last site of the lead interacting with the system stands $n$ sites away from it into the contact, then we can extend the system and include $n+1$ sites of the lead in it, thus, coming to the point interaction condition.

In the point interaction approximation functions $P$ and $Q$ reduce themselves to:
\begin{equation}
\begin{split}
P&=2\sqrt{\Gamma_{L}\Gamma_{R}}M_{N1},\\
Q&=\det{\left(\omega\hat I-\hat H_{aux}\right)},
\end{split}
\label{eq:PQpoint}
\end{equation}
where $M_{N1}$ is a minor of $(\omega\hat I-\hat H_{eff})$ matrix, $\hat H_{eff}$ is the effective Hamiltonian:
\begin{multline}
\left(\hat H_{eff}\right)_{mn}=\left(\hat H_{0}\right)_{mn}\\
+\left(\delta_{L}-i\Gamma_{L}\right)\delta_{m1}\delta_{n1}+\left(\delta_{R}-i\Gamma_{R}\right)\delta_{mN}\delta_{nN},
\label{eq:Heffpoint}
\end{multline}
and $\hat H_{aux}$ is the auxiliary Hamiltonian:
\begin{multline}
\left(\hat H_{aux}\right)_{mn}=\left(\hat H_{0}\right)_{mn}\\
+\left(\delta_{L}+i\Gamma_{L}\right)\delta_{m1}\delta_{n1}+\left(\delta_{R}-i\Gamma_{R}\right)\delta_{mN}\delta_{nN}.
\label{eq:Hauxpoint}
\end{multline}
The feature of Eq.~(\ref{eq:PQpoint}) is that minor $M_{N1}$ turns out to be independent of $\delta_{L(R)}$ and $\Gamma_{L(R)}$. Thus we can treat $M_{N1}$ here as a minor of the $(\omega\hat I-\hat H_{0})$ matrix. This approximation of point interaction with contacts will be used in Sec.~\ref{Sec.BIC} and Sec.~\ref{Sec.ColZer}.

\section{Bound states in the continuum}
\label{Sec.BIC}
\subsection{General properties of $P$ and $Q$ functions at BIC}
Bound state in the continuum (BIC) is a localized state with energy lying within energy interval of continuum states.\cite{bib:Hsu} BICs are non-decaying states, hence, they do not interact with continuum and, therefore, have zero width. Such states correspond to real eigenvalues of the effective Hamiltonian lying within the energy band of contacts. In Ref.~\cite{bib:QDR2,bib:QDR3,bib:ABQD1,bib:ABQD3} BIC in some particular QD systems were identified by the presence of a $\delta$-function peak in the density of states (DOS). This result can be generalized for an arbitrary two terminal system (see Appendix~\ref{ApG} for details). Here we discuss connection between BICs and properties of $P$ and $Q$ functions.

Suppose effective Hamiltonian $\hat H_{eff}$ has a real eigenvalue $\omega=\omega_{0}$: \begin{equation}
\det{\left(\omega\hat I-\hat H_{eff}\right)}\propto\left(\omega-\omega_{0}\right).
\label{eq:HeffBIC}
\end{equation}
According to Eq.~(\ref{eq:HeffPQ}) it follows from (\ref{eq:HeffBIC}) that:
\begin{equation}
\left|P\left(\omega\right)\right|^{2}+\left|Q\left(\omega\right)\right|^{2}=\left|\det{\left(\omega\hat I-\hat H_{eff}\right)}\right|^{2}\propto\left(\omega-\omega_{0}\right)^{2}.
\label{eq:PQBIC}
\end{equation}
At $\omega=\omega_{0}$ the sum of two non-negative quantities takes zero value, hence, they are both zero. Therefore:
\begin{equation}
P,Q\propto(\omega-\omega_{0}).
\label{eq:PQGCD}
\end{equation}
and we can conclude that there is a BIC in the system if and only if the $P$ and $Q$ share the same real root. 

These simple considerations, based on the introduction of the auxiliary Hamiltonian, show that the presence of a BIC at energy $\omega_{0}$ implies the presence of the root of $P$ at the same energy, which is responsible for transmission zeroes (\ref{eq:TransPQ}). Thus, according to Eqs.~(\ref{eq:TransHeff}) and~(\ref{eq:TransPQ}) the problem of divergence of transmission at BIC's energy, discussed, for example, in Ref.\cite{bib:BICDiverg}, is resolved easily. On the other hand, the reverse is {\it not} true and the presence of a real root of $P$ does not imply that there is a BIC. Shortly, one can formulate different possibilitiies as follows:
\begin{itemize}
\item there is a unity-valued resonance at energy $\omega_{0}$, if $P(\omega_{0})\neq0$ and $Q(\omega_{0})=0$;
\item there is a zero-valued antiresonance at energy $\omega_{0}$, if $P(\omega_{0})=0$ and $Q(\omega_{0})\neq0$;
\item there is a BIC at energy $\omega_{0}$, if $P(\omega_{0})=0$ and $Q(\omega_{0})=0$.
\end{itemize}
In more general case, suppose that energy $\omega_{0}$ is a root of multiplicity $m_{Q}$ of $Q$ and also it is a root of multiplicity $m_{P}$ of $P$. Then there are $\min{(m_{Q},m_{P})}$ degenerate BICs at energy $\omega_{0}$ and $m_{Q}-m_{P}$ coalesced resonances, if $m_{Q}>m_{P}$, or $m_{P}-m_{Q}$ coalesced antiresonances, if $m_{Q}<m_{P}$. If $m_{Q}=m_{P}$ then there are no extreme points of transmission at all. In the following subsection we illustrate this conclusion.

\subsection{Resonances and BICs in a toy three-site model}
\label{Sec.Exampl}
In this section we consider resonances, anti-resonances and BIC in a simple three-site structure (see inset in Fig.~\ref{fig6}a). The Hamiltonian of the structure has the form~(\ref{eq:HTot}) with the particular parameters defined below. We assume contacts to be the identical semi-infinite linear chains with equal on-site energies set as energy origin and nearest-neighbor hopping integrals $J$ set as energy unit. These contacts we treat to be attached locally to the site $0$ of the structure via equal tunneling matrix elements $\gamma^{L}_{0}=\gamma^{R}_{0}=\gamma\in\mathbb{R}$. Tunneling matrix elements $\tau_{a}$, $\tau_{b}$ and $\eta$ we also assume to be real and on-site energies of the structure we set to $\varepsilon_{0}=0$ and $\varepsilon_{a}=\varepsilon_{b}=\varepsilon$. Moreover, for the sake of convenience, we treat $\tau_{a}$ and $\tau_{b}$ to be of the same sign, e.g. positive. Thus, explicit expressions for the functions $P$ and $Q$ are
\begin{widetext}
\begin{equation}
\begin{split}
P&=\gamma^{2}\sqrt{4-\omega^{2}}\left(\omega-\varepsilon-\eta\right)\left(\omega-\varepsilon+\eta\right),\\
Q&=\left(\omega-\varepsilon+\eta\right)^{3}\left(1-\gamma^{2}\right)+\left(\omega-\varepsilon+\eta\right)^{2}\left(\varepsilon-3\eta\right)\left(1-\gamma^{2}\right)+\left(\omega-\varepsilon+\eta\right)\left[2\eta\left(\eta-\varepsilon\right)\left(1-\gamma^{2}\right)-\tau_{a}^{2}-\tau_{b}^{2}\right]\\
&+\eta\left(\tau_{a}-\tau_{b}\right)^{2}.
\end{split}
\label{eq:PBIC}
\end{equation}
\end{widetext}
$P$ has two real roots $\omega=\varepsilon\pm\eta$, which coincide with eigenstates of QD molecule formed by quantum dots $a$ and $b$. Another two roots $\omega^2=4$ correspond to contact band edges where particle velocity and hence the transparency turn into zero. BIC occurs if and only if $Q$ shares roots with $P$. According to Eq.~(\ref{eq:PQBIC}), only root $\omega=\varepsilon-\eta$ can be common for $P$ and $Q$ and this takes place as soon as $\tau_{a}=\tau_{b}$ or $\eta=0$. If $\tau_{a}$ and $\tau_{b}$ were of opposite signs, there would be a BIC at $\tau_{a}=-\tau_{b}$ and with energy $\omega=\varepsilon+\eta$. This BIC has properties similar to that at $\tau_{a}=\tau_{b}$ and with energy $\omega=\varepsilon-\eta$, thus, we will not consider it in this paper. Multiplicities $m_{P}$ and $m_{Q}$ of roots of $P$ and $Q$ can vary and in the following subsections we consider different cases of BIC formation, depending on these multiplicities. 

This structure possess conventional bound states with exponential decaying asymptotics: $a^{L(R)}_{n}\propto e^{-\kappa|n|}$. However, energies of these bound states lay out of a continuum band, i.e. they are bound states outside the continuum (BOC). On the other hand there can be a BIC, with all site amplitudes set to zero except for $a$ and $b$, which are related to each other as
\begin{align}
\label{eq:BICAmpls1}
&\tau_{a}a+\tau_{b}b=0,\quad for\quad\eta=0,\\
\label{eq:BICAmpls2}
&a=-b,\quad for\quad\tau_{a}=\tau_{b},\\
\label{eq:BICAmpls3}
\end{align}
and can both be non-zero. 
To find particular values for $a$ and $b$ in each case one can normalize it as a conventional bound state. 
As we see from (\ref{eq:BICAmpls1} and \ref{eq:BICAmpls3}) BIC amplitude distribution is antisymmetric for $\tau_{a}=\tau_{b}$. 

Scattering state site amplitudes in this structure can be easily calculated by solving corresponding tight-binding Schroedinger equation with boundary conditions given by plane-wave ansatz in contacts: $a^{L}_{n}=e^{ikn}+re^{-ikn}$ in the left one and $a^{R}_{n}=te^{ikn}$ in the right one, where $r$ and $t$ are reflection and transmission amplitudes respectively. Taking into account dispersion relation in contacts ($\omega=-2\cos{k}$) one can derive scattering state site amplitudes:
\begin{equation}
\begin{split}
 a_{0}&=C\left(\omega-\varepsilon-\eta\right)\left(\omega-\varepsilon+\eta\right),\\
a&=C\left[\tau_{a}\left(\omega-\varepsilon+\eta\right)-\eta\left(\tau_{a}-\tau_{b}\right)\right],\\
b&=C\left[\tau_{b}\left(\omega-\varepsilon+\eta\right)-\eta\left(\tau_{b}-\tau_{a}\right)\right],
\end{split}
\label{eq:ExSites}
\end{equation}
where
\begin{widetext}
\begin{equation*}
C=\frac{i\gamma\sqrt{4-\omega^{2}}}{\left(\omega-\varepsilon+\eta\right)\left\{\left(\omega-\varepsilon-\eta\right)\left[\omega-\gamma^{2}\left(\omega-i\sqrt{4-\omega^{2}}\right)\right]-\tau_{a}^{2}-\tau_{b}^{2}\right\}+\eta\left(\tau_{a}-\tau_{b}\right)^{2}}.
\end{equation*}
\end{widetext}
From~(\ref{eq:ExSites}) it follows that either at BIC or at antiresonance energy (both corresponding to $P=0$) site amplitude $a_{0}$ always equals zero.

\subsubsection{$m_{Q}<m_{P}$}
\begin{figure}
\includegraphics{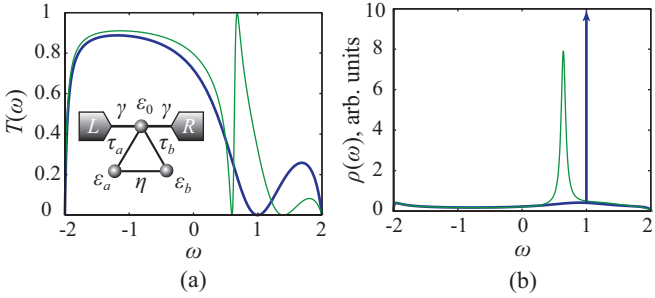}
\caption{\label{fig6}  (Color online) BIC at zero transmission energy. (a) Transmission coefficient of the three-site system within tight-binding approximation with $\eta=0.4$ (thin green line) and $\eta=0$ (thick blue line). In the inset: schematic view of the three-site structure considered. (b) DOS in this structure for the same parameters as in the part (a). At $\eta=0$ BIC appears and manifests itself as a $\delta$-function peak in the DOS. All values are in units of $J$.}
\end{figure}

For our certain structure, described above, condition $m_{Q}<m_{P}$ can be fulfilled only for $m_{Q}=1$ and $m_{P}=2$, which in turn requires  $\eta=0$. In this case BIC forms at the energy of zero transmittance $\omega=\varepsilon$. Figure~\ref{fig6} depicts plots of transmission coefficient and DOS for $\eta=0$ and for $\eta\neq0$. For illustration we take $\gamma=1$, $\tau_{a}=1$, $\tau_{b}=0.5$ and $\varepsilon=1$. According to Appendix~\ref{ApG}, formation of a BIC manifests itself as a $\delta$-function peak of DOS.

Now consider site amplitudes of the scattering state in the vicinity of this BIC. According to Eq.~(\ref{eq:ExSites}) at energy $\omega=\varepsilon$ there are no special features of the site amplitude distribution for any values of $\eta$ and even for $\eta=0$, corresponding to the BIC formation. Nevertheless, from the direct analysis of Eq.~(\ref{eq:ExSites}) one can figure out that energy $\omega=\varepsilon'$, where
\begin{equation}
\varepsilon'=\varepsilon-2\eta\frac{\tau_{a}\tau_{b}}{\tau_{a}^{2}+\tau_{b}^{2}}
\label{eq:Ene}
\end{equation}
scattering state site amplitudes vector $(a_{0},a,b)$ becomes
\begin{widetext}
\begin{equation}
\left(a_{0},a,b\right)=\frac{i\gamma\left(\tau_{a}^{2}+\tau_{b}^{2}\right)\sqrt{4-\varepsilon'^{2}}}{\left(\tau_{a}^{2}-\tau_{b}^{2}\right)\left[\varepsilon'\left(1-\gamma^{2}\right)+i\gamma^{2}\sqrt{4-\varepsilon'^{2}}\right]}\left(\frac{\tau_{a}^{2}-\tau_{b}^{2}}{\tau_{a}^{2}+\tau_{b}^{2}},\frac{\tau_{b}}{\eta},-\frac{\tau_{a}}{\eta}\right).
\label{eq:strBICam1}
\end{equation}
\end{widetext}
As one can see from Eq.~(\ref{eq:strBICam1}), scattering state amplitudes $a$ and $b$ are distributed in a full accordance with the BIC, i.e. satisfy the relation~(\ref{eq:BICAmpls1}), and, moreover, they formally diverge as $\eta$ tends to zero. On the other hand, at the exact BIC condition ($\eta=0$ and $\omega=\varepsilon$) amplitudes are
\begin{equation}
\left(a_{0},a,b\right)=-\frac{i\gamma\sqrt{4-\varepsilon^{2}}}{\tau_{a}^{2}+\tau_{b}^{2}}\left(0,\tau_{a},\tau_{b}\right).
\label{eq:strBICam2}
\end{equation}
From Eq.~(\ref{eq:strBICam2}) one can see that scattering state site amplitudes distribution ($a$ and $b$) at the exact BIC condition abruptly changes and becomes orthogonal to the BIC. For $\tau_{a}=\tau_{b}$ it corresponds to change from antisymmteric to symmetric state.

\subsubsection{$m_{Q}=m_{P}$}
\begin{figure}
\includegraphics{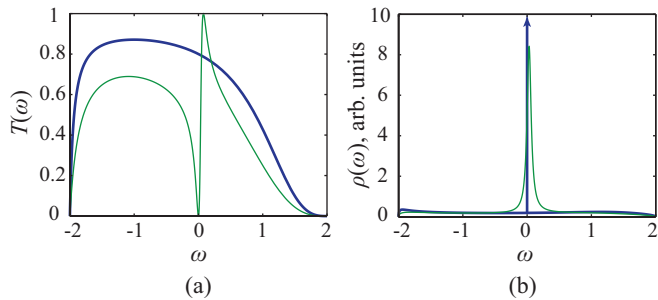}
\caption{\label{fig4} (Color online) BIC at a point of ``annihilation'' of a Fano resonance-antiresonance pair with non-zero transmission. (a) Transmission coefficient of the three-site system within tight-binding approximation with $\tau_{a}=1.5$ (thin green line) and $\tau_{a}=1$ (thick blue line). (b) DOS in this structure for the same parameters as in the part (a). At $\tau_{a}=1$ BIC appears and manifests itself as a $\delta$-function peak in the DOS. All values are in units of $J$.}
\end{figure}

When $m_{Q}=m_{P}$ BIC forms at the energy, corresponding to a non-extreme point of the transmission. For the structure we consider, the only possible case is $m_{Q}=m_{P}=1$. According to Eq.~(\ref{eq:PBIC}) this requires $P$ and $Q$ to be linear in $\left(\omega-\varepsilon+\eta\right)$, which can be satisfied if  $\tau_{a}=\tau_{b}$ and $\eta\neq0$. For instance, let us take $\eta=1\neq0$, $\gamma=1$, $\tau_{a}=\tau_{b}=1$ and $\varepsilon=1$. In this particular case BIC forms at energy $\omega=\varepsilon-\eta=0$. Figure~\ref{fig4} shows the transmission coefficient and the DOS of the three-site structure with $\tau_{a}=\tau_{b}$ and with $\tau_{a}\neq\tau_{b}$.

At the exact energy of this BIC ($\omega=\varepsilon-\eta$), as can be deduced from Eq.~(\ref{eq:ExSites}), scattering state amplitudes $(a_{0},a,b)$ are excited in an anti-symmetric way:
\begin{equation}
a_{0}=0,\quad a=-b=\frac{i\gamma\sqrt{4-\left(\varepsilon-\eta\right)^{2}}}{\tau_{b}-\tau_{a}}.
\label{eq:AmplA}
\end{equation}
This distribution fully coincides in symmetry with the corresponding BIC~(\ref{eq:BICAmpls2}). In the limit $\tau_{a}\rightarrow\tau_{b}$ amplitudes $a$ and $b$ formally diverge. However, in the exact BIC regime ($\tau_{a}=\tau_{b}$) and at energy $\omega=\varepsilon-\eta$, as can be seen from Eq.~(\ref{eq:ExSites}), scattering state amplitudes $(a_{0},a,b)$ are
\begin{widetext}
\begin{equation}
\left(a_{0},a,b\right)=-\frac{i\gamma\sqrt{4-\left(\varepsilon-\eta\right)^{2}}}{2\eta\left[\varepsilon-\eta-\gamma^{2}\left(\varepsilon-\eta-i\sqrt{4-\left(\varepsilon-\eta\right)^{2}}\right)\right]-2\tau_{a}^{2}}\left(0,\tau_{a},\tau_{a}\right).
\label{eq:AmplA2}
\end{equation}
\end{widetext}
According to Eq.~(\ref{eq:AmplA2}) $a$ and $b$ site amplitudes are distributed symmetrically and are orthogonal to the BIC.

\subsubsection{$m_{Q}>m_{P}$}

\begin{figure}
\includegraphics{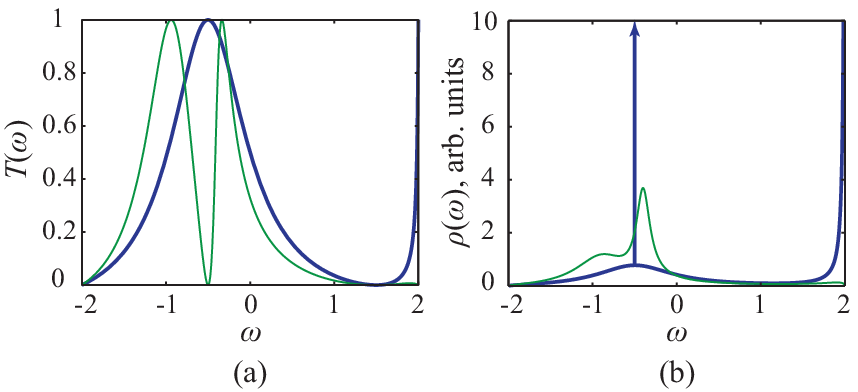}
\caption{\label{fig5} (Color online) BIC at perfect transmission energy. (a) Transmission coefficient of the three-site system within tight-binding approximation with $\tau_{a}=1$ (thin green line) and $\tau_{a}=\frac{\sqrt{3}}{2\sqrt{2}}$ (thick blue line). (b) DOS in this structure for the same parameters as in the part (a). At $\tau_{a}=\frac{\sqrt{3}}{2\sqrt{2}}$ BIC appears and manifests itself as a $\delta$-function peak in the DOS. Perfect transmission and Van Hove singularity of DOS in the upper band edge are due to the real root of $Q$ exact in this band edge. All values are in units of $J$.}
\end{figure}

For $m_{Q}>m_{P}$ the BIC forms at the energy of the perfect transmission. According to the particular structure we consider, this can take place only for $m_{Q}=2$ and $m_{P}=1$.  
In this case $P$ should be linear and $Q$ should be quadratic on $\left(\omega-\varepsilon+\eta\right)$. Thus from Eq.~(\ref{eq:PQBIC}) we deduce that $\eta\neq0$ and $\tau_{a}=\tau_{b}=\sqrt{\eta(\eta-\varepsilon)(1-\gamma^{2})}$. In order to have a non-disjoint structure ($\tau_{a}$ and $\tau_{b}$ cannot vanish simultaneously) we also should restrict ourselves with $\eta>\varepsilon$ and $\gamma<1$. As an example let us take $\eta=1\neq0$, $\varepsilon=0.5<\eta$ and $\gamma=0.5<1$, which results in $\tau_{a}=\tau_{b}=\frac{\sqrt{3}}{2\sqrt{2}}$. At these conditions BIC forms at energy $\omega=\varepsilon-\eta=-0.5$. Figure~\ref{fig5} illustrates this by plots of transmission coefficient and DOS at $\tau_{a}=\tau_{b}$ and at $\tau_{a}\neq\tau_{b}$. Site amplitudes distribution in this case does not differ from the case with $m_{P}=m_{Q}$ and is governed by Eqs.~(\ref{eq:AmplA}-\ref{eq:AmplA2}). Although, special parameters choice here leads to the perfect resonance formation in the BIC regime.

Yet another feature, which is specific to the particular choice of the parameters, shown in Fig.~\ref{fig5} is a perfect transmission at the upper band edge, where group velocity turns into zero. It results from Van Hove singularity of the DOS ( $\rho\sim\frac{1}{\sqrt{2-\omega}}$) and can be understood easily using properties of $P$ and $Q$ functions. Indeed, at the parameters corresponding to the BIC ($\tau_{a}=\tau_{b}$), one can see from Eq.~(\ref{eq:PBIC}) that polynomial $Q$ has a real root $\omega=2\eta$. For our choice $\eta=1$ we get that $Q$ has a real root in the very upper band edge. Thus, we have $Q\sim(2-\omega)$ and $P\sim\sqrt{2-\omega}$, which obviously results in the perfect transmission at $\omega=2$.
The phenomenon of perfect band edge transmission  is common for real roots of $Q$, falling at the very band edge with Van Hove singularity.

\subsection{BIC formation as a ``ghost phase transition'' with an abrupt symmetry transformation}
In the precceding section for the particular three-site toy model we obtained a singularity in the scattering state site amplitudes approaching BIC point in the parameter space. Near the BIC point symmetry of the scattering state amplitudes on sites forming BIC coincide with the symmetry of the BIC amplitudes. At the very BIC point symmetry of the scattering state amplitude abruptly changes. Here we show that this singularity, which is the manifestation of an abrupt symmetry transformation, is a general property of a system in the parameter space region near BIC state.
We consider a point contact approximation and also assume contacts to be semi-infinite linear chains with on-site energies set as energy origin and nearest neighbor hopping integral $J$ set as energy unit. Site amplitudes vector $\mathbf{a}=(a_{1},...,a_{N})^{\intercal}$ can be found from the following equation: 
\begin{equation}
\omega\hat I\mathbf{a}=\hat H_{eff}\mathbf{a}+\mathbf{s},
\label{eq:EqAmpl}
\end{equation}
where $\hat I$ is the $N\times N$ identity matrix and $\mathbf{s}$ is a source vector. Such a form of equation can be easily deduced, i.e., from the results of Ref.~\cite{bib:Jin2011}. In our case source vector is $\mathbf{s}=(s,0,...,0)^{\intercal}$ with
\begin{equation}
s=\frac{2i}{\gamma_{1}^{L}}\Gamma_{L}.
\label{eq:s}
\end{equation}
From Eq.~(\ref{eq:EqAmpl}) we can straightforwardly find sites amplitudes:
\begin{equation}
\mathbf{a}=\left(\omega\hat I-\hat H_{eff}\right)^{-1}\mathbf{s}=\left(\hat A+i\mathbf{u}_{L}\mathbf{u}_{L}^{\dag}+i\mathbf{u}_{R}\mathbf{u}_{R}^{\dag}\right)^{-1}\mathbf{s}.
\label{eq:Ampl}
\end{equation}
Under the assumption of point interaction each of the vectors $\mathbf{u}_{L(R)}$ in the site localized states has only one nonzero element $u_{L,i}=\delta_{i1}\sqrt{\Gamma_{L}}$ and $u_{R,i}=\delta_{RN}\sqrt{\Gamma_{R}}$ correspondingly.

Next we transform the basis to hybridized eigenstates of the structure, which diagonalize the matrix $\hat A$. For a sake of definiteness, we suppose that BIC originates from the state $\ket{\tilde 1}$. It takes place as soon as couplings $\tilde u_{L(R),\tilde 1}$ vanish. Here tilde highlights the eigenstate basis. In the vicinity of this BIC one can approximate amplitude $\tilde a_{\tilde 1}$ of the $\ket{\tilde 1}$ state as
\begin{equation}
\tilde a_{\tilde 1}\approx\frac{\alpha\left(\sqrt{\tilde\Gamma_{L}},\sqrt{\tilde\Gamma_{R}}\right)}{\Delta\omega+\beta\left(\sqrt{\tilde\Gamma_{L}},\sqrt{\tilde\Gamma_{R}}\right)},
\label{eq:BICAmpl}
\end{equation}
where $\Delta\omega=\omega-\tilde\varepsilon_{\tilde 1}$ with $\tilde\varepsilon_{\tilde 1}$ being the energy of the $\ket{\tilde 1}$ state. In Eq.~(\ref{eq:BICAmpl}) $\alpha(x,y)$ is a some linear form of $x$ and $y$, $\beta(x,y)$ is a some bilinear form of $x$ and $y$ and $\tilde\Gamma_{L(R)}=|\tilde u_{L(R),\tilde 1}|^{2}$. It should be noted here that the state $\ket{\tilde 1}$ and, consequently, its energy $\tilde\varepsilon_{\tilde 1}=\tilde\varepsilon_{\tilde 1}(\tilde u_{L(R),\tilde 1})$ depend on parameters on their own and, thus, it becomes a BIC just in the limit $\tilde u_{L(R),\tilde 1}\rightarrow0$ (or $\tilde\Gamma_{L(R)}\rightarrow0$).

\begin{figure}
\includegraphics{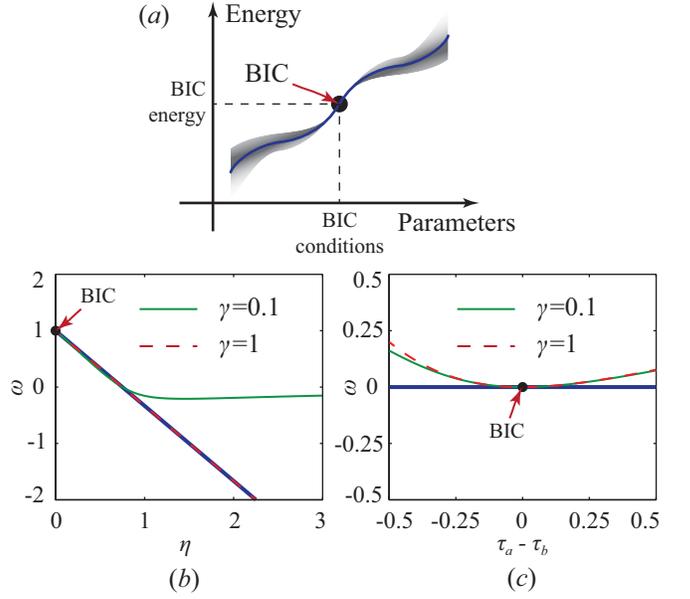}
\caption{\label{fig9} (Color online) Trajectories demonstarating scattering state amplitude divergence in the energy-parameters space. (a) General view of a trajectory of $\tilde\varepsilon_{\tilde 1}(\tilde u_{L(R),\tilde 1})$ (blue line) and a schematic region (shaded), where trajectories $\omega(\tilde u_{L(R),\tilde 1})$ providing scattering state amplitude  formal divergence can pass. (b) Exact hybridized eigenenergies of the three-site structure from the previous subsection in the vicinity of the BIC at $\eta=0$ and $\omega=\varepsilon$ with following parameters $\tau_{a}=1$, $\tau_{b}=0.5$, $\varepsilon=1$ and $\gamma=0.1$ (thin solid green line) or $\gamma=1$ (thin dashed red line). Thick blue line corresponds to the trajectory $\omega=\varepsilon'$. It is easily seen that this trajectory fulfill the condition~(\ref{eq:DerCond}). (c) Exact hybridized eigenenergies of the same structure in the vicinity of the BIC at $\tau_{a}=\tau_{b}$ and $\omega=\varepsilon-\eta$ with $\tau_{b}=1$, $\varepsilon=1$, $\eta=1$ and $\gamma=0.1$ (thin solid green line) or $\gamma=1$ (thin dashed red line). Thick blue line corresponds to the trajectory $\omega=\varepsilon-\eta$. In this case trajectory providing the formal divergence is simple (constant) because derivative of $\tilde\varepsilon_{\tilde 1}$ in the exact BIC is zero. All values are in units of $J$.}
\end{figure}

From the Eq.~(\ref{eq:BICAmpl}) one can conclude that if one approaches BIC energy by a some trajectory in the energy-parameters space $\omega=\omega(\tilde u_{L(R),\tilde 1})$ such that
\begin{equation}
\left.\frac{\partial\omega\left(\tilde u_{L(R),\tilde 1}\right)}{\partial\tilde u_{L(R),\tilde 1}}\right|_{\tilde u_{L(R),\tilde 1}=0}=\left.\frac{\partial\tilde\varepsilon_{\tilde 1}\left(\tilde u_{L(R),\tilde 1}\right)}{\partial\tilde u_{L(R),\tilde 1}}\right|_{\tilde u_{L(R),\tilde 1}=0},
\label{eq:DerCond}
\end{equation}
then $\Delta\omega=\omega-\tilde\varepsilon_{\tilde 1}=\mathcal{O}(|\tilde u_{L(R),\tilde 1}|^{2})=\mathcal{O}(\tilde\Gamma_{L(R)})$ and the amplitude of the scattering state $\tilde a_{\tilde 1}$ formally diverges with parameters tending to the BIC condition ($\tilde u_{L(R),\tilde 1},\tilde\Gamma_{L(R)}\rightarrow0$). Figure~\ref{fig9}a schematically illustrates this concept. This general description sheds light on the features of the scattering state site amplitudes behavior in the example considered in details in the previous subsection. It turned out that for the BIC at $\eta=0$ and $\omega=\varepsilon$ trajectory, which provides the formal divergence of the scattering state amplitudes, is given by $\omega=\varepsilon'$ with $\varepsilon'$ from Eq.~(\ref{eq:Ene}), while for the BIC at $\tau_{a}=\tau_{b}$ and $\omega=\varepsilon-\eta$ diverging trajectory is simple (constant independent on $\tau_{a(b)}$): $\omega=\varepsilon-\eta$. Both this trajectories fulfill the condition~(\ref{eq:DerCond}), which is illustrated by Fig.~\ref{fig9}b and Fig.~\ref{fig9}c respectively.


On the other hand, if parameters are at the exact BIC condition ($\tilde u_{L(R),\tilde 1}=0$), then amplitude $\tilde a_{\tilde 1}$ identically equals to zero with a removable singularity at $\Delta\omega=0$ ($\omega=\tilde\varepsilon_{\tilde 1}$). Therefore, from the analysis of the Eq.~(\ref{eq:BICAmpl}), we get that the scattering state amplitude distribution in the vicinity of the BIC corresponds to the distribution of $\ket{\tilde1}$ state (the same as in for the BIC) and it abruptly changes to orthogonal one (such that $\tilde a_{\tilde1}=0$), if the exact BIC condition is fulfilled. Thus, BIC formation can be understood in some sense as a ``phase transition'' resulting in abrupt symmetry transformation. In terms of Landau theory of phase transitions incident wave can be considered as conjugated field to ``order parameter'', which is described by BIC amplitude, and  diverging state amplitude $\tilde a_{\tilde1}$ corresponds to the Curie-Weiss response function near the phase transition point.

\section{Engineering Fano resonances: coalescence of resonances and antiresonances}
\label{Sec.ColZer}
\subsection{Quantum dot loop: generalized meta-coupling with leads}
\begin{figure}
\includegraphics{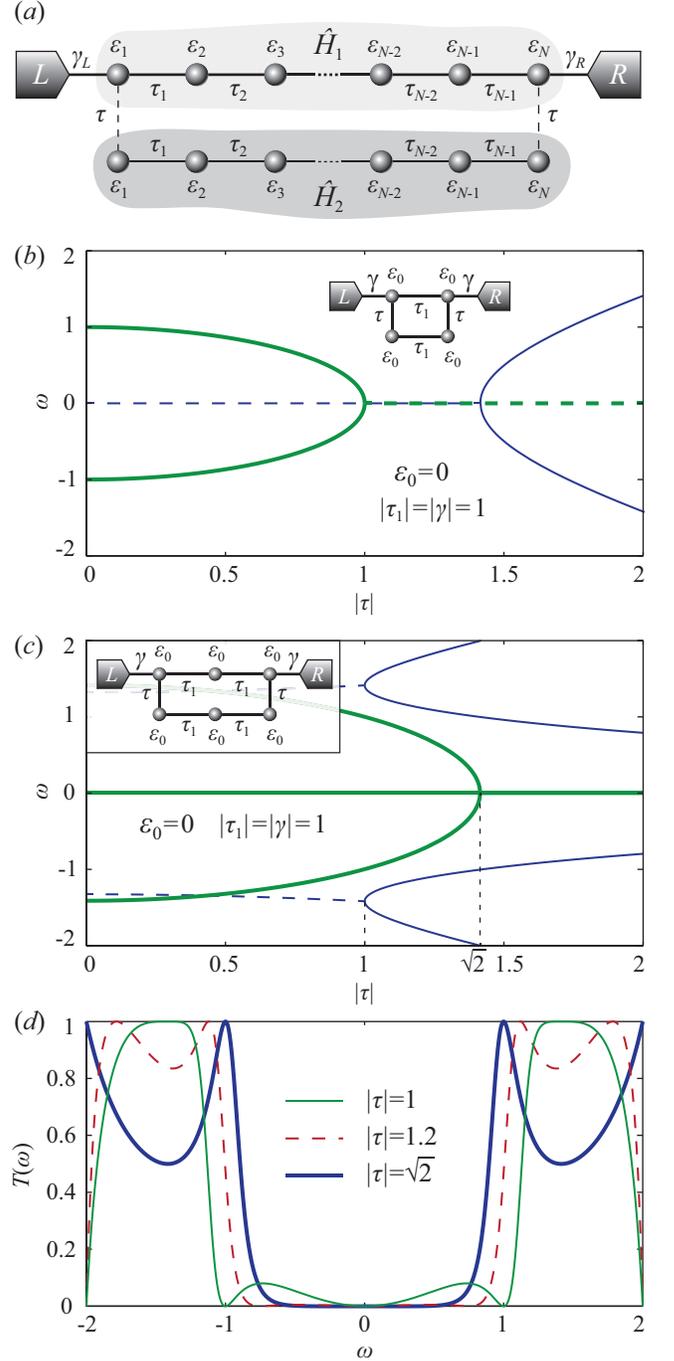}
\caption{\label{fig7} (Color online) Coalescence of antiresonances in a double chain structure. (a) Schematic view of a double chain structure. (b, c) Real parts of the roots of the $P(\omega)$ (thick green line) and real parts of roots of the $Q(\omega)$ (thin blue line) calculated for the $N=2$-site double chain (b), which is shown in the inset, and for the $N=3$-site double chain (c), which is also shown in the inset. Solid lines show entirely real roots and dashed lines stand for real parts of complex roots. (d) Transmission coefficient profile in the very regime of antiresonance coalescence (thick blue line), in the very regime of resonance coalescence (thin green line) and in an intermediate regime (thin dashed red line). All values are in units of $J$.}
\end{figure}
Coalescence of perfect transmission maximums was shown Ref.\cite{bib:GorShJETP,bib:GorShAop} to occur at the EP of the non-Hermitian auxiliary Hamiltonian. Here we focus on the coalescence of transmission zeros (antiresonances) and show that it can be related to an EP of some additional non-Hermitian Hamiltonian as well. We consider the structure (Fig.~\ref{fig7}a), consisting of two equal $N$-site chains connected to each other by tunneling matrix element $\tau$ between the edge sites. We numerate sites from $1$ to $N$ in each chain, thus, contacts are connected to the first and to the $N$-th site of the first chain via matrix elements $\gamma_{L}$ and $\gamma_{R}$ correspondingly. Hence the number of sites in these chains forming two branches of the loop differ by two and such a coupling can be considered as a generalization of the meta-coupling widely studied in aromatic molecules\cite{bib:Aroma2}. The Hamiltonian of the system we consider is of general form~(\ref{eq:HTot}) with a special choice of on-site energies $\varepsilon_{i}$ and hopping integrals $\tau_{ij}$. Bare Hamiltonian $\hat H_{0}$ of this structure is more convenient to write in a block form:
\begin{equation}
\begin{split}
\hat H_{0}&=\hat H_{1}+\hat H_{2}+\left(\hat\Omega_{12}+h.c.\right),\\
\hat H_{1}&=\sum_{i=1}^{N}{\varepsilon_{i}a_{i}^{\dag}a_{i}}+\sum_{i=1}^{N-1}{\left(\tau_{i}a_{i+1}^{\dag}a_{i}+h.c.\right)},\\
\hat H_{2}&=\sum_{i=1}^{N}{\varepsilon_{i}b_{i}^{\dag}b_{i}}+\sum_{i=1}^{N-1}{\left(\tau_{i}b_{i+1}^{\dag}b_{i}+h.c.\right)},\\
\hat\Omega_{12}&=\tau b_{1}^{\dag}a_{1}+\tau b_{N}^{\dag}a_{N},
\label{eq:HamColZer}
\end{split}
\end{equation}
where $a_{i}^{\dag}(a_{i})$ and $b_{i}^{\dag}(b_{i})$ are creation (annihilation) operators in the $i$-th site of the first and the second chains respectively. Here $\hat H_{1(2)}$ corresponds to the Hamiltonian of the first (second) chain and $\hat\Omega_{12}$ describes interaction between them.

Now we assume that the system is symmetric ($\tau_{i}=\tau_{N-i}$ and $\gamma_{L}=\gamma_{R}=\gamma$) and has identical on-site energies: ($\varepsilon_{i}=\varepsilon_{0}$ for each $i$)  Using Eq.~(\ref{eq:PQpoint}) one can calculate function $P$ for this case (see Appendix \ref{ApB} for details):
\begin{equation}
P=2\sqrt{\Gamma_{L}\Gamma_{R}}\tau_{1}\cdot...\cdot\tau_{N-1}\det{\left(\omega\hat I-\hat H_{zero}\right)}.
\label{eq:PColZer}
\end{equation}
Here $\hat H_{zero}$ is a $\mathcal{PT}$-symmetric Hamiltonian defined in~(\ref{eq:Hzero}), of which real eigenvalues determine transmission zeroes just as eigenvalues of auxiliary Hamiltonian~(\ref{eq:Haux}). Thus, for chains invariant under mirror reflection and having identical on-site energies there can take place a coalescence of $N$ zeros of transmission as an $N$-th order EP of the Hamiltonian $\hat H_{zero}$.  Hence according to Ref.\cite{bib:GorShAop}  coalescence of even number of transmission zeros results in a nonzero dip, whereas coalescence of odd number of zeros results in a zero-valued dip. Transmission coefficient, according to the general relation~(\ref{eq:TransPQ}), near a real $N$-th order root $\omega_{0}$ of polynomial $P$ takes form:
\begin{equation}
T(\omega)=\frac{\left(\omega-\omega_{0}\right)^{2N}}{\left(\omega-\omega_{0}\right)^{2N}+\tilde\Gamma^{2N}},
\label{eq:TransColZer}
\end{equation}
where $\tilde\Gamma$ is some energy-dependent parameter and $\tilde\Gamma(\omega_{0})\neq0$.

As an illustration we consider an $N=2$-site double chain and $N=3$-site double chain structures. Contacts in both cases are treated as semi-infinite linear chains with hoping integral $J$ set as energy unit. Figure~\ref{fig7}b shows real roots and real parts of complex roots of polynomial $P$ and $Q$ for the $2$-site double chain as functions of $|\tau|$. We set $|\tau_{1}|=|\gamma|=1$ and $\varepsilon_{0}=0$. Figure~\ref{fig7}c corresponds to the $3$-site double chain. Here we again assume $|\tau_{1}|=|\gamma|=1$ and $\varepsilon_{0}=0$. Coalescence of real roots of polynomial $P$ (shown by thick green lines in Fig.~\ref{fig7}b and Fig.~\ref{fig7}c) indeed corresponds to the coalescence of transmission zeros, because it takes place at the nonzero point of polynomial $Q$. For the particular examples considered it is not difficult to derive conditions for the coalescence of antiresonances: $|\tau|=1$ for the 2-site double chain structure (Fig.~\ref{fig7}b) and $|\tau|=\sqrt{2}$ for the 3-site double chain structure (Fig.~\ref{fig7}c). These plots also demonstrate difference between coalescence of even and odd number of antiresonances mentioned above. Figure~\ref{fig7}d shows transmission vs. energy profiles for the 3-site double chain structure. These profiles are plotted for three values of $|\tau|$ representing the coalescence of antiresonances ($|\tau|=\sqrt{2}$), coalescence of two pair of resonances ($|\tau|=1$) and in some intermediate position ($|\tau|=1.2$). For $|\tau|=\sqrt{2}$ here are perfect transmission points on the band edges, which are due to the real roots of $Q$, located exactly at the band edges (see Fig.~\ref{fig7}c).

\subsection{Quantum dot comb structure: crossing of antiresonances}
\label{Sec.CoalFano}
\begin{figure}
\includegraphics{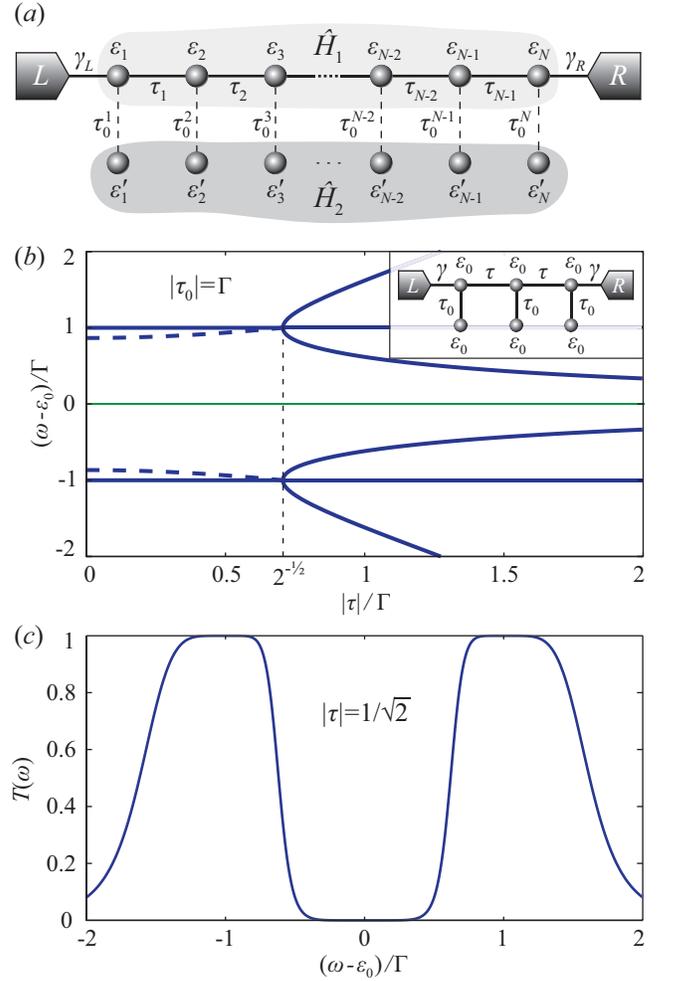}
\caption{\label{fig-1} (Color online) Coalescence of Fano resonances and antiresonances in a comb-like structure. (a) Schematic view of a comb-like structure. (b) Positions of the real parts of the roots of the $P(\omega)$ (thin green line) and real parts of roots of the $Q(\omega)$ (thick blue line) calculated for the $N=3$-site comb-like structure, which is shown in the inset, in the wide band limit. Solid lines show entirely real roots and dashed lines stand for real parts of complex roots. Parameter $|\tau_{0}|$ is set to $1$. One can see that coalescence of resonances takes place at $|\tau|=\frac{1}{\sqrt{2}}\Gamma$. (c) Transmission coefficient profile in the very regime of resonance coalescence.}
\end{figure}

Consider a comb-like structure representing an $N$-site linear chain with side-defect sites connected to each site of the chain (Fig.~\ref{fig-1}a). The bare Hamiltonian of this structure is also more convenient to write in a block form:
\begin{equation}
\begin{split}
\hat H_{0}&=\hat H_{1}+\hat H_{2}+\left(\hat\Omega_{12}+h.c.\right),\\
\hat H_{1}&=\sum_{i=1}^{N}{\varepsilon_{i}a_{i}^{\dag}a_{i}}+\sum_{i=1}^{N-1}{\left(\tau_{i}a_{i+1}^{\dag}a_{i}+h.c.\right)},\\
\hat H_{2}&=\sum_{i=1}^{N}{\varepsilon_{i}'b_{i}^{\dag}b_{i}},\\
\hat\Omega_{12}&=\sum_{i=1}^{N}{\tau_{0}^{i} b_{i}^{\dag}a_{i}}.
\end{split}
\label{eq:HamComb}
\end{equation}
Here $a_{i}^{\dag}(a_{i})$ is a creation (annihilation) operator in the $i$-th site of the chain with energy $\varepsilon_{i}$ and $b_{i}^{\dag}(b_{i})$ is a creation (annihilation) operator in the $i$-th side-defect site with energy $\varepsilon_{i}'$ connected to the $i$-th site of the chain via hopping integral $\tau_{0}^{i}$.

We assume that all sites of the linear chain and all side-defect sites are physically identical, i.e. have the same energy: $\varepsilon_{i}'=\varepsilon_{i}=\varepsilon_{0}$, also we suppose that $\tau_{0}^{1}=...=\tau_{0}^{N}=\tau_{0}$. Contacts are treated as identical and are connected to the $1$-st and to the $N$-th site of the chain by matrix elements $\gamma_{L}=\gamma_{R}=\gamma$ (resulting in $\Gamma_{L}=\Gamma_{R}=\Gamma$ and $\delta_{L}=\delta_{R}=\delta$). In this case, functions $P$ and $Q$ can be derived in the following form (see Appendix~\ref{ApE} for details):
\begin{widetext}
\begin{equation}
P=2\tilde\omega^{N}\Gamma\tau_{1}\cdot...\cdot\tau_{N-1},\qquad
Q=\begin{vmatrix}
\left(\tilde\omega-\delta-i\Gamma\right)\tilde\omega-\left|\tau_{0}\right|^{2} & -\tau_{1}\tilde\omega & \ldots & 0 & 0\\
-\tau_{1}^{*}\tilde\omega & \tilde\omega^{2}-\left|\tau_{0}\right|^{2} & \ldots & 0 & 0\\
\vdots & \vdots & \ddots & \vdots & \vdots\\
0 & 0 & \ldots & \tilde\omega^{2}-\left|\tau_{0}\right|^{2} & -\tau_{N-1}\tilde\omega\\
0 & 0 & \ldots & -\tau_{N-1}^{*}\tilde\omega & \left(\tilde\omega-\delta+i\Gamma\right)\tilde\omega-\left|\tau_{0}\right|^{2}
\end{vmatrix},
\label{eq:PQComb}
\end{equation}
\end{widetext}
where $\tilde\omega=\omega-\varepsilon_{0}$.

From Eq.~(\ref{eq:PQComb}) it is clear that $\omega=\varepsilon_{0}$ ($\tilde\omega=0$) is an $N$-th order root of $P$ and $Q(\tilde\omega=0)\neq0$, hence $\omega=\varepsilon_{0}$ is an $N$-th order zero of transmission. This is a crossing point of Fano resonance minima. In the wide band limit (or Fermi golden rule approximation)\cite{bib:Pastawski2008} there can be also a coalescence of Fano resonance maxima in this structure as well. Under this assumption $\delta\approx0$ and $\Gamma\approx\gamma^{2}/J\approx const$ is independent of energy. Here $J$ is a half of the band width in the contacts, which we treat to be much greater than difference between energies of our interest ($\omega$, $\varepsilon_{0}$) and contacts band center. According to the Eq.~(\ref{eq:PQComb}), coalescence of transmission peaks can take place at the energy $\omega=\varepsilon_{0}\pm|\tau_{0}|$ at a certain ratio between tunneling matrix elements $\tau_{i}$.\cite{bib:GorShJETP,bib:GorShAop} Thus, one can tune the parameters of the structure in such a way, that its transmission coefficient will have an $N$-th order zero dip surrounded by two $N$-th order unity peaks. Figure~\ref{fig-1}b shows positions of real parts of the roots of the functions $Q$ and $P$ for the $3$-site comb-like structure calculated in the wide band limit. Coalescence of $3$ real roots of $Q$, forms an EP of the $3$-rd order, which corresponds to the coalescence of resonances. Figure~\ref{fig-1}c depicts related profile of transmission coefficient energy dependence in the very regime of coalescence of resonances.

\section{Summary}
\label{Sec.Sum}
In this paper we have presented a general description of resonances, antiresonances and BICs via the unique formalism. Our observation is that dissipationless but open quantum system possesses features such as EP and $\mathcal{PT}$-symmetry breaking, which are common to systems with balanced gain and loss terms in the Hamiltonian.  We showed that for an arbitrary two terminal multiple connected molecular (or QD) conductor or waveguide the square module of the characteristic determinant of the effective Hamiltonian (denominator in the expression for the transparency) can be written in a simple form of a sum of two non-negative terms. The first term is a square module of the characteristic determinant of the auxiliary Hamiltonian, which zeroes determine the transparency peaks. The second term is an energy (frequency) dependent function that is exactly the numerator in the expression for the transparency and its zeroes determine antiresonances. The non-Hermitian auxiliary Hamiltonian can be easily deduced from the Feshbach's effective non-Hermitian Hamiltonian. Resonances and BICs are related to complex and real eigenvalues of effective Hamiltonian correspondingly. However, a complex eigenvalue of the effective Hamiltonian, which is a pole of the scattering matrix, in general, doesn't determine the position of the resonance on the energy axis. Real eigenvalues of the auxiliary Hamiltonian which coincide with real eigenvalues of effective Hamiltonian determine BICs. Real eigenvalues of the auxiliary Hamiltonian, which don't coincide with real eigenvalues of effective Hamiltonian, determine the exact positions of perfect resonances on the energy axis. EPs of the auxiliary Hamiltonian are responsible for the coalescence of resonances. It should be noted also that in this paper all calculations were carried out within a localized orthogonal basis (constructed, for example, by L\"{o}wdin orthogonalization~\cite{bib:Lowdin,*bib:Lowdin1962,*bib:Lowdin1963}). On the other hand, in numerical simulations of real quantum molecular conductors (e.g. in DFT) the basis of Hamiltonian eigenstates, which diagonalizes the initial Hamiltonian of the isolated structure, is more convenient. It can be shown that in diagonal basis antiresonances are described by nondiagonal non-Hermitian coupling in the effective Hamiltonian.\cite{bib:Thomas,bib:QDR2,bib:ABQD1,bib:FanoEP,bib:Longhi2009}

Scattering states and BICs are deeply coupled to each other. As we have shown, symmetry (mutual interrelation) of the scattering state amplitudes on the sites corresponding to the BIC exactly coincides with the symmetry of the BIC amplitudes near the BIC point in the parameter space. A trajectory in the generalized energy-parameter space can be chosen such that on this trajectory the absolute value of the scattering state amplitude diverges while approaching the BIC point. At the very BIC point structure of the scattering state amplitudes changes abruptly and the scattering state wave function (waveguide mode) becomes orthogonal to the BIC. This picture closely resembles behavior of a system near the second order phase transition point with the BIC being the order parameter, scattering state wave function being the conjugated field and scattering state amplitudes at the BIC sites -- the generalized response function obeying Curie-Weiss-like law. Our model doesn't account for the interelectron Coulomb interactions, which can be partially justified under the assumption of a strong coupling with contacts resulting in small values of site amplitudes inside the molecule. However, for electromagnetic fields in waveguides the description is adequate for large site amplitudes as well. Hence, our model provides a straightforward approach for creating a BIC and a storage of intense fields by abrupt switching from the scattering regime to the BIC and vice versa.

Obtained results, which relate resonance and BIC energies to the problem of finding real roots of a well-defined energy functions, make it possible to control positions of perfect and zero transmission as well as their coalescence. Thus, our results could be helpful for the deduction of the design rules for quantum conductors and waveguides. For example, one can convert the perfect transmission into the zero transmission (or vice versa) at the same energy, by tuning some structure's parameters. As an example of such design rules application we have constructed two families of quantum structures -- asymmetric loop with symmetric branches (generalized meta coupling to the leads) and symmetric comb structure, which exhibit coalescence of antiresonance resulting in the formation of a broad reflection window. In the former structure an almost rectangle window of transparency, described in\cite{bib:GorShAop}, is converted into an almost rectangle window of reflectivity~(\ref{eq:TransColZer}) just by adding the same single-channel wire in parallel. Such rectangle windows of transmission can be applied\cite{bib:RectTrans} within the area of quantum heat engines.\cite{bib:QuantHeat,bib:QHEPRB} Other fields of possible applications are broad band filters design\cite{bib:Filt2,*bib:Filt3} and 2D photonic crystals\cite{bib:PhotCrNat,bib:PhotCrFano} formed by 2D periodic arrays of dielectric rods with an in-plane light wave, which is polarized along these rods.\cite{bib:Kivsh1,*bib:Kivsh2,*bib:Mir1,*bib:TransZeros} In particular, one can get the transmission dips or peaks by adding or removing defects close to the main waveguide. Moreover, one can create both transmission window or reflection window using single-chain and double-chain structures correspondingly. Thus, with the same lithographic template both structures can be realized.

\begin{acknowledgments}
AAG would like to acknowledge the Program of Fundamental Research of the Presidium of the Russian Academy of Sciences for partial support. 
\end{acknowledgments}

\appendix

\section{Density of states in the vicinity of a BIC}
\label{ApG}
Here we show that for an arbitrary QD system described by the Hamiltonian (\ref{eq:HTot}) formation of a BIC results in appearance of $\delta$-like peak of density of states (DOS). DOS can be derived straightforwardly from the retarded Green's function of the system~(\ref{eq:GFHeff}):
\begin{equation}
\rho(\omega)=-\frac{1}{\pi}\I{\sum_{i}{G^{r}_{ii}}}=-\frac{1}{\pi}\I{\Tr{\hat G^{r}}},
\label{eq:DOSdef}
\end{equation}
where summation runs through all sites of the structure. In terms of matrix $\hat A$ and vectors $\mathbf{u}_{L(R)}$, introduced in Sec.~\ref{Sec.Trans}, the DOS can be written as:
\begin{equation}
\rho(\omega)=-\frac{1}{\pi}\I{\Tr{\left(\hat A+i\mathbf{u}_{L}\mathbf{u}_{L}^{\dag}+i\mathbf{u}_{R}\mathbf{u}_{R}^{\dag}\right)^{-1}}}.
\label{eq:DOSA}
\end{equation}
Using Sherman–Morrison formula\cite{bib:SherMorr} and matrix determinant lemma\cite{bib:MatDetLem} one can get an explicit expression for the DOS in the following form:
\begin{equation}
\rho(\omega)=\frac{1}{\pi}\frac{R(\omega)}{\left|P(\omega)\right|^2+\left|Q(\omega)\right|^2}
\label{eq:DOS}
\end{equation}
with $P$ and $Q$ are defined by Eqs.~(\ref{eq:P}) and~(\ref{eq:Q}) and
\begin{widetext}
\begin{multline}
R=\left(\det{\hat A}\right)^{2}\left\{\mathbf{u}_{L}^{\dag}\hat A^{-2}\mathbf{u}_{L}\left[1+\left(\mathbf{u}_{R}^{\dag}\hat A^{-1}\mathbf{u}_{R}\right)^{2}\right]+\mathbf{u}_{R}^{\dag}\hat A^{-2}\mathbf{u}_{R}\left[1+\left(\mathbf{u}_{L}^{\dag}\hat A^{-1}\mathbf{u}_{L}\right)^{2}\right]\right.\\
\left.-2\left(\mathbf{u}_{L}^{\dag}\hat A^{-1}\mathbf{u}_{L}+\mathbf{u}_{R}^{\dag}\hat A^{-1}\mathbf{u}_{R}\right)\R{\left[\mathbf{u}_{R}^{\dag}\hat A^{-1}\mathbf{u}_{L}\mathbf{u}_{L}^{\dag}\hat A^{-2}\mathbf{u}_{R}\right]}+\left|\mathbf{u}_{R}^{\dag}\hat A^{-1}\mathbf{u}_{L}\right|^{2}\left(\mathbf{u}_{L}^{\dag}\hat A^{-2}\mathbf{u}_{L}+\mathbf{u}_{R}^{\dag}\hat A^{-2}\mathbf{u}_{R}\right)\right\}.
\label{eq:R}
\end{multline}
\end{widetext}

BIC is a localized state, which is totally decoupled from the continuum of states in the  leads. Thus, in the basis of eigenstates of the structure, hybridized by the leads (i.e. in the basis, which diagonalizes matrix $\hat A$), BIC can be understood as a state $\ket{\tilde n}$, which couplings to the leads $\tilde \gamma_{\tilde n}^{L(R)}$ and, hence, corresponding ($n$-th) element of the vector $\mathbf{\tilde u}_{L(R)}$ vanish.\cite{bib:ABQD1} Here tilde highlights the diagonalized basis.

Now consider DOS in the vicinity of a BIC. Suppose that $\tilde\varepsilon_{\tilde 1},...,\tilde\varepsilon_{\tilde N}$ are eigenenergies of the hybridized structure, and  we assume for definiteness that BIC appears at energy $\omega=\tilde\varepsilon_{\tilde 1}$, when parameters $\tilde u_{L(R),\tilde 1}$ tend to zero. In this case matrix $\hat A$ is diagonal with $A_{\tilde i\tilde i}=(\omega-\tilde\varepsilon_{\tilde i})$. In the vicinity of the BIC we can assume $\Delta\omega=\omega-\tilde\varepsilon_{\tilde 1}$ and $\tilde\Gamma_{L(R)}=|\tilde u_{L(R),\tilde 1}|^{2}$ to be small compared to $\min_{\tilde i, \tilde j}{|\tilde\varepsilon_{\tilde i}-\tilde\varepsilon_{\tilde j}|}$ and $\min_{\tilde i}{|\tilde u_{L(R),\tilde i}|^{2}}$ correspondingly. Treating $\tilde \Gamma_{L(R)}$ and $\Delta\omega$ as small quantities of the same order, one can approximate $\rho(\omega)$ in the vicinity of a BIC as:
\begin{widetext}
\begin{equation}
\rho(\omega)\approx\frac{1}{\pi}\frac{\beta_{1}\left(\sqrt{\tilde\Gamma_{L}},\sqrt{\tilde\Gamma_{R}}\right)}{\left[\Delta\omega+\beta_{2}\left(\sqrt{\tilde\Gamma_{L}},\sqrt{\tilde\Gamma_{R}}\right)\right]^2+\left[\beta_{3}\left(\sqrt{\tilde\Gamma_{L}},\sqrt{\tilde\Gamma_{R}}\right)\right]^2},
\label{eq:DOSBIC}
\end{equation}
\end{widetext}
where $\beta_{i}(a,b)$ are some bilinear forms of $a$ and $b$. From Eq.~(\ref{eq:DOSBIC}) it is clear that in the limit $\tilde\Gamma_{L(R)}\rightarrow0$ DOS in the vicinity of a BIC has a $\delta$-function peak.

\section{Derivation of function $P$ for a double chain structure}
\label{ApB}
To calculate the function $P$ for a double chain structure we use Eq.~(\ref{eq:PQpoint}), where minor $M_{1N}$ of $(\omega\hat I-\hat H_{eff})$ is needed. According to Eq.~(\ref{eq:HamColZer}), the Hamiltonian of the isolated system $\hat H_{0}$ and, consequently, the effective Hamiltonian $\hat H_{eff}$ can be presented as a $2\times 2$ block matrix:
\begin{equation}
\hat H_{eff}=\begin{pmatrix}
\hat H_{1}^{eff} & \hat\Omega_{12}\\
\hat\Omega_{12}^{\dag} & \hat H_{2}
\end{pmatrix}.
\label{eq:BlockHeff}
\end{equation}
Here $\hat H_{1}^{eff}$ is a Hamiltonian of the first chain with contact self-energies taken into account. Applying the rules of block matrix inversion\cite{bib:BlMatInv} (or equivalently L\"{o}wdin partitioning technique\cite{bib:Lowdin,*bib:Lowdin1962,*bib:Lowdin1963}) one can calculate necessary minor and get the following expression for the function $P$:
\begin{widetext}
\begin{equation}
P=2\sqrt{\Gamma_{L}\Gamma_{R}}\det{\left(\omega\hat I-\hat H_{2}\right)}\cdot\begin{vmatrix}
             -\tau_{1} & 0 & 0 & \ldots & 0 & 0 & -|\tau|^{2}\left[\left(\omega\hat I-\hat H_{2}\right)^{-1}\right]_{1N}\\
             \omega & -\tau_{2} & 0 & \ldots & 0 & 0 & 0\\
             -\tau_{2}^{*} & \omega & -\tau_{3} & \ldots & 0 & 0 & 0\\
             \vdots & \vdots & \vdots & \ddots & \vdots & \vdots & \vdots\\
             0 & 0 & 0 & \ldots & -\tau_{N-3} & 0 & 0\\
             0 & 0 & 0 & \ldots & \omega & -\tau_{N-2} & 0\\
             0 & 0 & 0 & \ldots & -\tau_{N-2}^{*} & \omega & -\tau_{N-1}
             \end{vmatrix}.
\label{eq:B1}
\end{equation}
Matrix element $[(\omega\hat I-\hat H_{2})^{-1}]_{1N}$ is derived from Eq.~(\ref{eq:HamColZer}):
\begin{equation}
\left[\left(\omega\hat I-\hat H_{2}\right)^{-1}\right]_{1N}=\frac{\tau_{1}\cdot...\cdot\tau_{N-1}}{\det{\left(\omega\hat I-\hat H_{2}\right)}}.
\label{eq:B2}
\end{equation}
Substituting~(\ref{eq:B2}) into~(\ref{eq:B1}) and expanding determinant by the first row, one can simplify $P$ in the following way:
\begin{multline}
P=2\sqrt{\Gamma_{L}\Gamma_{R}}\det{\left(\omega\hat I-\hat H_{2}\right)}\left[(-1)^{N-1}\tau_{1}\cdot...\cdot\tau_{N-1}-(-1)^{N}|\tau|^{2}\frac{\tau_{1}\cdot...\cdot\tau_{N-1}}{\det{\left(\omega\hat I-\hat H_{2}\right)}}D_{1}^{1}\right]\\
=2\sqrt{\Gamma_{L}\Gamma_{R}}\cdot(-1)^{N-1}\tau_{1}\cdot...\cdot\tau_{N-1}\left[\det{\left(\omega\hat I-\hat H_{2}\right)}+|\tau|^{2}D_{1}^{1}\right].
\label{eq:B3}
\end{multline}
\end{widetext}
Here $D_{q}^{p}$ stands for the minor of the $(\omega\hat I-\hat H_{1})$ matrix with the first $p$ rows and columns and the last $q$ rows and columns crossed out. As chains are equal we can also think of a $D_{q}^{p}$ as the corresponding minor of the $(\omega\hat I-\hat H_{2})$ matrix.

Now we use the fact that the two-chain structure under study is symmetric. In this case we can conclude that the expression in square brackets in Eq.~(\ref{eq:B3}) is a determinant of the matrix $(\omega\hat I-\hat H_{zero})$, where $\hat H_{zero}$ is the following $\mathcal{PT}$-symmetric Hamiltonian:
\begin{equation}
\left(\hat H_{zero}\right)_{mn}=\left(\hat H_{2}\right)_{mn}+i|\tau|\left(\delta_{m1}\delta_{n1}-\delta_{mN}\delta_{nN}\right).
\label{eq:Hzero}
\end{equation}
Indeed, this can be checked directly by expanding the determinant of $(\omega\hat I-\hat H_{zero})$:
\begin{multline}
\det{\left(\omega\hat I-\hat H_{zero}\right)}=\det{\left(\omega\hat I-\hat H_{2}\right)}+i\left|\tau\right|\left(D_{0}^{1}-D_{1}^{0}\right)\\
+\left|\tau\right|^{2}D_{1}^{1}.
\label{eq:B4}
\end{multline}
In symmetric structure minors $D_{0}^{1}$ and $D_{1}^{0}$ are equal and from Eq.~(\ref{eq:B4}) we get exactly the expression in square brackets in the r.h.s. of Eq.~(\ref{eq:B3}).

As was mentioned in the Sec.~\ref{Sec.Trans}, $P$ is defined up to an arbitrary phase factor. Thus, we can neglect the sign in the Eq.~(\ref{eq:B3}) and get the polynomial $P$ for the symmetric two-chain model in the form~(\ref{eq:PColZer}).

\section{Derivation of functions $P$ and $Q$ for a comb-like structure}
\label{ApE}
As it was done for a double chain structure, in the case of a comb-like structure, we can again write the effective Hamiltonian in the block form~(\ref{eq:BlockHeff}), but with $\hat H_{1}$, $\hat H_{2}$ and $\hat\Omega_{12}$ taken from Eq.~(\ref{eq:HamComb}). Such a form of the effective Hamiltonian allows us to calculate $P$ easily:
\begin{equation}
P=2\sqrt{\Gamma_{L}\Gamma_{R}}\tau_{1}\cdot...\cdot\tau_{N-1}\left(\omega-\varepsilon_{1}'\right)\cdot...\cdot\left(\omega-\varepsilon_{N}'\right).
\label{eq:PComb}
\end{equation}
Function $Q$ can be derived in a similar way as $P$, because the auxiliary Hamiltonian also has a block matrix form. Thus, according to Eq.~(\ref{eq:PQpoint}) and once again using block matrix inversion rules from Ref.\cite{bib:BlMatInv} one can get that
\begin{widetext}
\begin{multline}
Q=\begin{vmatrix}
\omega-\varepsilon_{1}-\delta_{L}-i\Gamma_{L}-\frac{\left|\tau_{0}^{1}\right|^{2}}{\omega-\varepsilon_{1}'} & -\tau_{1} & \ldots & 0 & 0\\
-\tau_{1}^{*} & \omega-\varepsilon_{2}-\frac{\left|\tau_{0}^{2}\right|^{2}}{\omega-\varepsilon_{2}'} & \ldots & 0 & 0\\
\vdots & \vdots & \ddots & \vdots & \vdots\\
0 & 0 & \ldots & \omega-\varepsilon_{N-1}-\frac{\left|\tau_{0}^{N-1}\right|^{2}}{\omega-\varepsilon_{N-1}'} & -\tau_{N-1}\\
0 & 0 & \ldots & -\tau_{N-1}^{*} & \omega-\varepsilon_{N}-\delta_{R}+i\Gamma_{R}-\frac{\left|\tau_{0}^{N}\right|^{2}}{\omega-\varepsilon_{N}'}
\end{vmatrix}\\
\times\left(\omega-\varepsilon_{1}'\right)\cdot...\cdot\left(\omega-\varepsilon_{N}'\right).
\label{eq:QComb}
\end{multline}
\end{widetext}
Assuming that $\varepsilon_{i}'=\varepsilon_{i}=\varepsilon_{0}$, $\tau_{0}^{1}=...=\tau_{0}^{N}=\tau_{0}$ and contacts are identical, we simplify Eqs.~(\ref{eq:PComb}-\ref{eq:QComb}) to the Eq.~(\ref{eq:PQComb}).

\bibliography{references}

\begin{thebibliography}{90}%
\makeatletter
\providecommand \@ifxundefined [1]{%
 \@ifx{#1\undefined}
}%
\providecommand \@ifnum [1]{%
 \ifnum #1\expandafter \@firstoftwo
 \else \expandafter \@secondoftwo
 \fi
}%
\providecommand \@ifx [1]{%
 \ifx #1\expandafter \@firstoftwo
 \else \expandafter \@secondoftwo
 \fi
}%
\providecommand \natexlab [1]{#1}%
\providecommand \enquote  [1]{``#1''}%
\providecommand \bibnamefont  [1]{#1}%
\providecommand \bibfnamefont [1]{#1}%
\providecommand \citenamefont [1]{#1}%
\providecommand \href@noop [0]{\@secondoftwo}%
\providecommand \href [0]{\begingroup \@sanitize@url \@href}%
\providecommand \@href[1]{\@@startlink{#1}\@@href}%
\providecommand \@@href[1]{\endgroup#1\@@endlink}%
\providecommand \@sanitize@url [0]{\catcode `\\12\catcode `\$12\catcode
  `\&12\catcode `\#12\catcode `\^12\catcode `\_12\catcode `\%12\relax}%
\providecommand \@@startlink[1]{}%
\providecommand \@@endlink[0]{}%
\providecommand \url  [0]{\begingroup\@sanitize@url \@url }%
\providecommand \@url [1]{\endgroup\@href {#1}{\urlprefix }}%
\providecommand \urlprefix  [0]{URL }%
\providecommand \Eprint [0]{\href }%
\providecommand \doibase [0]{http://dx.doi.org/}%
\providecommand \selectlanguage [0]{\@gobble}%
\providecommand \bibinfo  [0]{\@secondoftwo}%
\providecommand \bibfield  [0]{\@secondoftwo}%
\providecommand \translation [1]{[#1]}%
\providecommand \BibitemOpen [0]{}%
\providecommand \bibitemStop [0]{}%
\providecommand \bibitemNoStop [0]{.\EOS\space}%
\providecommand \EOS [0]{\spacefactor3000\relax}%
\providecommand \BibitemShut  [1]{\csname bibitem#1\endcsname}%
\let\auto@bib@innerbib\@empty
\bibitem [{\citenamefont {Moiseyev}(2011)}]{bib:MoiseyevBook}%
  \BibitemOpen
  \bibfield  {author} {\bibinfo {author} {\bibfnamefont {N.}~\bibnamefont
  {Moiseyev}},\ }\href {\doibase 10.1017/CBO9780511976186} {\emph {\bibinfo
  {title} {Non-Hermitian Quantum Mechanics}}}\ (\bibinfo  {publisher}
  {Cambridge University Press},\ \bibinfo {year} {2011})\BibitemShut {NoStop}%
\bibitem [{\citenamefont {Monticone}\ and\ \citenamefont
  {Alù}(2017)}]{bib:Monti2017}%
  \BibitemOpen
  \bibfield  {author} {\bibinfo {author} {\bibfnamefont {F.}~\bibnamefont
  {Monticone}}\ and\ \bibinfo {author} {\bibfnamefont {A.}~\bibnamefont
  {Alù}},\ }\href {http://stacks.iop.org/0034-4885/80/i=3/a=036401} {\bibfield
   {journal} {\bibinfo  {journal} {Reports on Progress in Physics}\ }\textbf
  {\bibinfo {volume} {80}},\ \bibinfo {pages} {036401} (\bibinfo {year}
  {2017})}\BibitemShut {NoStop}%
\bibitem [{\citenamefont {Miroshnichenko}\ \emph {et~al.}(2010)\citenamefont
  {Miroshnichenko}, \citenamefont {Flach},\ and\ \citenamefont
  {Kivshar}}]{bib:FanoRev}%
  \BibitemOpen
  \bibfield  {author} {\bibinfo {author} {\bibfnamefont {A.~E.}\ \bibnamefont
  {Miroshnichenko}}, \bibinfo {author} {\bibfnamefont {S.}~\bibnamefont
  {Flach}}, \ and\ \bibinfo {author} {\bibfnamefont {Y.~S.}\ \bibnamefont
  {Kivshar}},\ }\href {\doibase 10.1103/RevModPhys.82.2257} {\bibfield
  {journal} {\bibinfo  {journal} {Rev. Mod. Phys.}\ }\textbf {\bibinfo {volume}
  {82}},\ \bibinfo {pages} {2257} (\bibinfo {year} {2010})}\BibitemShut
  {NoStop}%
\bibitem [{\citenamefont {Hsu}\ \emph {et~al.}(2016)\citenamefont {Hsu},
  \citenamefont {Zhen}, \citenamefont {Stone}, \citenamefont {Joannopoulos},\
  and\ \citenamefont {Solja\v{c}i\'{c}}}]{bib:Hsu}%
  \BibitemOpen
  \bibfield  {author} {\bibinfo {author} {\bibfnamefont {C.~W.}\ \bibnamefont
  {Hsu}}, \bibinfo {author} {\bibfnamefont {B.}~\bibnamefont {Zhen}}, \bibinfo
  {author} {\bibfnamefont {A.~D.}\ \bibnamefont {Stone}}, \bibinfo {author}
  {\bibfnamefont {J.~D.}\ \bibnamefont {Joannopoulos}}, \ and\ \bibinfo
  {author} {\bibfnamefont {M.}~\bibnamefont {Solja\v{c}i\'{c}}},\ }\href
  {http://dx.doi.org/10.1038/natrevmats.2016.48} {\ \textbf {\bibinfo {volume}
  {1}},\ \bibinfo {pages} {16048 EP } (\bibinfo {year} {2016})},\ \bibinfo
  {note} {review Article}\BibitemShut {NoStop}%
\bibitem [{\citenamefont {Landau}\ and\ \citenamefont
  {Lifshitz}(2013)}]{bib:LandauQM}%
  \BibitemOpen
  \bibfield  {author} {\bibinfo {author} {\bibfnamefont {L.~D.}\ \bibnamefont
  {Landau}}\ and\ \bibinfo {author} {\bibfnamefont {E.~M.}\ \bibnamefont
  {Lifshitz}},\ }\href@noop {} {\emph {\bibinfo {title} {Quantum mechanics:
  non-relativistic theory}}},\ Vol.~\bibinfo {volume} {3}\ (\bibinfo
  {publisher} {Elsevier},\ \bibinfo {year} {2013})\BibitemShut {NoStop}%
\bibitem [{\citenamefont {Kubala}\ and\ \citenamefont
  {K\"onig}(2002)}]{bib:QDR1}%
  \BibitemOpen
  \bibfield  {author} {\bibinfo {author} {\bibfnamefont {B.}~\bibnamefont
  {Kubala}}\ and\ \bibinfo {author} {\bibfnamefont {J.}~\bibnamefont
  {K\"onig}},\ }\href {\doibase 10.1103/PhysRevB.65.245301} {\bibfield
  {journal} {\bibinfo  {journal} {Phys. Rev. B}\ }\textbf {\bibinfo {volume}
  {65}},\ \bibinfo {pages} {245301} (\bibinfo {year} {2002})}\BibitemShut
  {NoStop}%
\bibitem [{\citenamefont {Guevara}\ \emph {et~al.}(2003)\citenamefont
  {Guevara}, \citenamefont {Claro},\ and\ \citenamefont {Orellana}}]{bib:QDR2}%
  \BibitemOpen
  \bibfield  {author} {\bibinfo {author} {\bibfnamefont {M.~L. L.~d.}\
  \bibnamefont {Guevara}}, \bibinfo {author} {\bibfnamefont {F.}~\bibnamefont
  {Claro}}, \ and\ \bibinfo {author} {\bibfnamefont {P.~A.}\ \bibnamefont
  {Orellana}},\ }\href {\doibase 10.1103/PhysRevB.67.195335} {\bibfield
  {journal} {\bibinfo  {journal} {Phys. Rev. B}\ }\textbf {\bibinfo {volume}
  {67}},\ \bibinfo {pages} {195335} (\bibinfo {year} {2003})}\BibitemShut
  {NoStop}%
\bibitem [{\citenamefont {Orellana}\ \emph {et~al.}(2004)\citenamefont
  {Orellana}, \citenamefont {Ladr\'on~de Guevara},\ and\ \citenamefont
  {Claro}}]{bib:QDR3}%
  \BibitemOpen
  \bibfield  {author} {\bibinfo {author} {\bibfnamefont {P.~A.}\ \bibnamefont
  {Orellana}}, \bibinfo {author} {\bibfnamefont {M.~L.}\ \bibnamefont
  {Ladr\'on~de Guevara}}, \ and\ \bibinfo {author} {\bibfnamefont
  {F.}~\bibnamefont {Claro}},\ }\href {\doibase 10.1103/PhysRevB.70.233315}
  {\bibfield  {journal} {\bibinfo  {journal} {Phys. Rev. B}\ }\textbf {\bibinfo
  {volume} {70}},\ \bibinfo {pages} {233315} (\bibinfo {year}
  {2004})}\BibitemShut {NoStop}%
\bibitem [{\citenamefont {Von~Neuman}\ and\ \citenamefont
  {Wigner}(1929)}]{bib:BIC1929}%
  \BibitemOpen
  \bibfield  {author} {\bibinfo {author} {\bibfnamefont {J.}~\bibnamefont
  {Von~Neuman}}\ and\ \bibinfo {author} {\bibfnamefont {E.}~\bibnamefont
  {Wigner}},\ }\href@noop {} {\bibfield  {journal} {\bibinfo  {journal}
  {Physikalische Zeitschrift}\ }\textbf {\bibinfo {volume} {30}},\ \bibinfo
  {pages} {467} (\bibinfo {year} {1929})}\BibitemShut {NoStop}%
\bibitem [{\citenamefont {Friedrich}\ and\ \citenamefont
  {Wintgen}(1985)}]{bib:Wintgen1985}%
  \BibitemOpen
  \bibfield  {author} {\bibinfo {author} {\bibfnamefont {H.}~\bibnamefont
  {Friedrich}}\ and\ \bibinfo {author} {\bibfnamefont {D.}~\bibnamefont
  {Wintgen}},\ }\href {\doibase 10.1103/PhysRevA.32.3231} {\bibfield  {journal}
  {\bibinfo  {journal} {Phys. Rev. A}\ }\textbf {\bibinfo {volume} {32}},\
  \bibinfo {pages} {3231} (\bibinfo {year} {1985})}\BibitemShut {NoStop}%
\bibitem [{\citenamefont {Kodigala}\ \emph {et~al.}(2017)\citenamefont
  {Kodigala}, \citenamefont {Lepetit}, \citenamefont {Gu}, \citenamefont
  {Bahari}, \citenamefont {Fainman},\ and\ \citenamefont
  {Kant{\'e}}}]{bib:BICLaser}%
  \BibitemOpen
  \bibfield  {author} {\bibinfo {author} {\bibfnamefont {A.}~\bibnamefont
  {Kodigala}}, \bibinfo {author} {\bibfnamefont {T.}~\bibnamefont {Lepetit}},
  \bibinfo {author} {\bibfnamefont {Q.}~\bibnamefont {Gu}}, \bibinfo {author}
  {\bibfnamefont {B.}~\bibnamefont {Bahari}}, \bibinfo {author} {\bibfnamefont
  {Y.}~\bibnamefont {Fainman}}, \ and\ \bibinfo {author} {\bibfnamefont
  {B.}~\bibnamefont {Kant{\'e}}},\ }\href
  {http://dx.doi.org/10.1038/nature20799} {\bibfield  {journal} {\bibinfo
  {journal} {Nature}\ }\textbf {\bibinfo {volume} {541}},\ \bibinfo {pages}
  {196} (\bibinfo {year} {2017})},\ \bibinfo {note} {letter}\BibitemShut
  {NoStop}%
\bibitem [{\citenamefont {Feshbach}(1958)}]{bib:Fesh1}%
  \BibitemOpen
  \bibfield  {author} {\bibinfo {author} {\bibfnamefont {H.}~\bibnamefont
  {Feshbach}},\ }\href {\doibase
  http://dx.doi.org/10.1016/0003-4916(58)90007-1} {\bibfield  {journal}
  {\bibinfo  {journal} {Annals of Physics}\ }\textbf {\bibinfo {volume} {5}},\
  \bibinfo {pages} {357 } (\bibinfo {year} {1958})}\BibitemShut {NoStop}%
\bibitem [{\citenamefont {Feshbach}(1962)}]{bib:Fesh2}%
  \BibitemOpen
  \bibfield  {author} {\bibinfo {author} {\bibfnamefont {H.}~\bibnamefont
  {Feshbach}},\ }\href {\doibase
  http://dx.doi.org/10.1016/0003-4916(62)90221-X} {\bibfield  {journal}
  {\bibinfo  {journal} {Annals of Physics}\ }\textbf {\bibinfo {volume} {19}},\
  \bibinfo {pages} {287 } (\bibinfo {year} {1962})}\BibitemShut {NoStop}%
\bibitem [{\citenamefont {Feshbach}(1967)}]{bib:Fesh3}%
  \BibitemOpen
  \bibfield  {author} {\bibinfo {author} {\bibfnamefont {H.}~\bibnamefont
  {Feshbach}},\ }\href {\doibase
  http://dx.doi.org/10.1016/0003-4916(67)90163-7} {\bibfield  {journal}
  {\bibinfo  {journal} {Annals of Physics}\ }\textbf {\bibinfo {volume} {43}},\
  \bibinfo {pages} {410 } (\bibinfo {year} {1967})}\BibitemShut {NoStop}%
\bibitem [{\citenamefont {Sasada}\ \emph {et~al.}(2011)\citenamefont {Sasada},
  \citenamefont {Hatano},\ and\ \citenamefont {Ordonez}}]{bib:Sasada2011}%
  \BibitemOpen
  \bibfield  {author} {\bibinfo {author} {\bibfnamefont {K.}~\bibnamefont
  {Sasada}}, \bibinfo {author} {\bibfnamefont {N.}~\bibnamefont {Hatano}}, \
  and\ \bibinfo {author} {\bibfnamefont {G.}~\bibnamefont {Ordonez}},\ }\href
  {\doibase 10.1143/JPSJ.80.104707} {\bibfield  {journal} {\bibinfo  {journal}
  {Journal of the Physical Society of Japan}\ }\textbf {\bibinfo {volume}
  {80}},\ \bibinfo {pages} {104707} (\bibinfo {year} {2011})}\BibitemShut
  {NoStop}%
\bibitem [{\citenamefont {Gorbatsevich}\ \emph {et~al.}(2008)\citenamefont
  {Gorbatsevich}, \citenamefont {Zhuravlev},\ and\ \citenamefont
  {Kapaev}}]{bib:Gor}%
  \BibitemOpen
  \bibfield  {author} {\bibinfo {author} {\bibfnamefont {A.}~\bibnamefont
  {Gorbatsevich}}, \bibinfo {author} {\bibfnamefont {M.}~\bibnamefont
  {Zhuravlev}}, \ and\ \bibinfo {author} {\bibfnamefont {V.}~\bibnamefont
  {Kapaev}},\ }\href {\doibase 10.1134/S106377610808013X} {\bibfield  {journal}
  {\bibinfo  {journal} {Journal of Experimental and Theoretical Physics}\
  }\textbf {\bibinfo {volume} {107}},\ \bibinfo {pages} {288} (\bibinfo {year}
  {2008})}\BibitemShut {NoStop}%
\bibitem [{\citenamefont {Vanroose}(2001)}]{bib:DoublePole2}%
  \BibitemOpen
  \bibfield  {author} {\bibinfo {author} {\bibfnamefont {W.}~\bibnamefont
  {Vanroose}},\ }\href {\doibase 10.1103/PhysRevA.64.062708} {\bibfield
  {journal} {\bibinfo  {journal} {Phys. Rev. A}\ }\textbf {\bibinfo {volume}
  {64}},\ \bibinfo {pages} {062708} (\bibinfo {year} {2001})}\BibitemShut
  {NoStop}%
\bibitem [{\citenamefont {Belozerova}\ and\ \citenamefont
  {Henner}(1998)}]{bib:Belozerova}%
  \BibitemOpen
  \bibfield  {author} {\bibinfo {author} {\bibfnamefont {T.}~\bibnamefont
  {Belozerova}}\ and\ \bibinfo {author} {\bibfnamefont {V.}~\bibnamefont
  {Henner}},\ }\href@noop {} {\bibfield  {journal} {\bibinfo  {journal}
  {Physics of Particles and Nuclei}\ }\textbf {\bibinfo {volume} {29}},\
  \bibinfo {pages} {63} (\bibinfo {year} {1998})}\BibitemShut {NoStop}%
\bibitem [{\citenamefont {Dente}\ \emph {et~al.}(2008)\citenamefont {Dente},
  \citenamefont {Bustos-Mar\'un},\ and\ \citenamefont
  {Pastawski}}]{bib:Pastawski2008}%
  \BibitemOpen
  \bibfield  {author} {\bibinfo {author} {\bibfnamefont {A.~D.}\ \bibnamefont
  {Dente}}, \bibinfo {author} {\bibfnamefont {R.~A.}\ \bibnamefont
  {Bustos-Mar\'un}}, \ and\ \bibinfo {author} {\bibfnamefont {H.~M.}\
  \bibnamefont {Pastawski}},\ }\href {\doibase 10.1103/PhysRevA.78.062116}
  {\bibfield  {journal} {\bibinfo  {journal} {Phys. Rev. A}\ }\textbf {\bibinfo
  {volume} {78}},\ \bibinfo {pages} {062116} (\bibinfo {year}
  {2008})}\BibitemShut {NoStop}%
\bibitem [{\citenamefont {Romo}\ and\ \citenamefont
  {Garc\'{\i}a-Calder\'on}(1994)}]{bib:Romo}%
  \BibitemOpen
  \bibfield  {author} {\bibinfo {author} {\bibfnamefont {R.}~\bibnamefont
  {Romo}}\ and\ \bibinfo {author} {\bibfnamefont {G.}~\bibnamefont
  {Garc\'{\i}a-Calder\'on}},\ }\href {\doibase 10.1103/PhysRevB.49.14016}
  {\bibfield  {journal} {\bibinfo  {journal} {Phys. Rev. B}\ }\textbf {\bibinfo
  {volume} {49}},\ \bibinfo {pages} {14016} (\bibinfo {year}
  {1994})}\BibitemShut {NoStop}%
\bibitem [{\citenamefont {M\"uller}\ \emph {et~al.}(1995)\citenamefont
  {M\"uller}, \citenamefont {Dittes}, \citenamefont {Iskra},\ and\
  \citenamefont {Rotter}}]{bib:rotter1995}%
  \BibitemOpen
  \bibfield  {author} {\bibinfo {author} {\bibfnamefont {M.}~\bibnamefont
  {M\"uller}}, \bibinfo {author} {\bibfnamefont {F.-M.}\ \bibnamefont
  {Dittes}}, \bibinfo {author} {\bibfnamefont {W.}~\bibnamefont {Iskra}}, \
  and\ \bibinfo {author} {\bibfnamefont {I.}~\bibnamefont {Rotter}},\ }\href
  {\doibase 10.1103/PhysRevE.52.5961} {\bibfield  {journal} {\bibinfo
  {journal} {Phys. Rev. E}\ }\textbf {\bibinfo {volume} {52}},\ \bibinfo
  {pages} {5961} (\bibinfo {year} {1995})}\BibitemShut {NoStop}%
\bibitem [{\citenamefont {Gorbatsevich}\ and\ \citenamefont
  {Shubin}(2017)}]{bib:GorShAop}%
  \BibitemOpen
  \bibfield  {author} {\bibinfo {author} {\bibfnamefont {A.}~\bibnamefont
  {Gorbatsevich}}\ and\ \bibinfo {author} {\bibfnamefont {N.}~\bibnamefont
  {Shubin}},\ }\href {\doibase http://dx.doi.org/10.1016/j.aop.2016.12.019}
  {\bibfield  {journal} {\bibinfo  {journal} {Annals of Physics}\ }\textbf
  {\bibinfo {volume} {376}},\ \bibinfo {pages} {353 } (\bibinfo {year}
  {2017})}\BibitemShut {NoStop}%
\bibitem [{\citenamefont {Khitrova}\ \emph {et~al.}(2006)\citenamefont
  {Khitrova}, \citenamefont {Gibbs}, \citenamefont {Kira}, \citenamefont
  {Koch},\ and\ \citenamefont {Scherer}}]{bib:Collapse1}%
  \BibitemOpen
  \bibfield  {author} {\bibinfo {author} {\bibfnamefont {G.}~\bibnamefont
  {Khitrova}}, \bibinfo {author} {\bibfnamefont {H.~M.}\ \bibnamefont {Gibbs}},
  \bibinfo {author} {\bibfnamefont {M.}~\bibnamefont {Kira}}, \bibinfo {author}
  {\bibfnamefont {S.~W.}\ \bibnamefont {Koch}}, \ and\ \bibinfo {author}
  {\bibfnamefont {A.}~\bibnamefont {Scherer}},\ }\href {\doibase
  10.1038/nphys227} {\bibfield  {journal} {\bibinfo  {journal} {Nat Phys}\
  }\textbf {\bibinfo {volume} {2}},\ \bibinfo {pages} {81} (\bibinfo {year}
  {2006})}\BibitemShut {NoStop}%
\bibitem [{\citenamefont {Stehmann}\ \emph {et~al.}(2004)\citenamefont
  {Stehmann}, \citenamefont {Heiss},\ and\ \citenamefont
  {Scholtz}}]{bib:EPelect}%
  \BibitemOpen
  \bibfield  {author} {\bibinfo {author} {\bibfnamefont {T.}~\bibnamefont
  {Stehmann}}, \bibinfo {author} {\bibfnamefont {W.~D.}\ \bibnamefont {Heiss}},
  \ and\ \bibinfo {author} {\bibfnamefont {F.~G.}\ \bibnamefont {Scholtz}},\
  }\href {http://stacks.iop.org/0305-4470/37/i=31/a=012} {\bibfield  {journal}
  {\bibinfo  {journal} {Journal of Physics A: Mathematical and General}\
  }\textbf {\bibinfo {volume} {37}},\ \bibinfo {pages} {7813} (\bibinfo {year}
  {2004})}\BibitemShut {NoStop}%
\bibitem [{\citenamefont {Schindler}\ \emph {et~al.}(2011)\citenamefont
  {Schindler}, \citenamefont {Li}, \citenamefont {Zheng}, \citenamefont
  {Ellis},\ and\ \citenamefont {Kottos}}]{bib:EPelect2}%
  \BibitemOpen
  \bibfield  {author} {\bibinfo {author} {\bibfnamefont {J.}~\bibnamefont
  {Schindler}}, \bibinfo {author} {\bibfnamefont {A.}~\bibnamefont {Li}},
  \bibinfo {author} {\bibfnamefont {M.~C.}\ \bibnamefont {Zheng}}, \bibinfo
  {author} {\bibfnamefont {F.~M.}\ \bibnamefont {Ellis}}, \ and\ \bibinfo
  {author} {\bibfnamefont {T.}~\bibnamefont {Kottos}},\ }\href {\doibase
  10.1103/PhysRevA.84.040101} {\bibfield  {journal} {\bibinfo  {journal} {Phys.
  Rev. A}\ }\textbf {\bibinfo {volume} {84}},\ \bibinfo {pages} {040101}
  (\bibinfo {year} {2011})}\BibitemShut {NoStop}%
\bibitem [{\citenamefont {Cardamone}\ \emph {et~al.}(2002)\citenamefont
  {Cardamone}, \citenamefont {Stafford},\ and\ \citenamefont
  {Barrett}}]{bib:EPQD1}%
  \BibitemOpen
  \bibfield  {author} {\bibinfo {author} {\bibfnamefont {D.}~\bibnamefont
  {Cardamone}}, \bibinfo {author} {\bibfnamefont {C.}~\bibnamefont {Stafford}},
  \ and\ \bibinfo {author} {\bibfnamefont {B.}~\bibnamefont {Barrett}},\ }\href
  {\doibase 10.1002/1521-3951(200204)230:2<419::AID-PSSB419>3.0.CO;2-I}
  {\bibfield  {journal} {\bibinfo  {journal} {physica status solidi (b)}\
  }\textbf {\bibinfo {volume} {230}},\ \bibinfo {pages} {419} (\bibinfo {year}
  {2002})}\BibitemShut {NoStop}%
\bibitem [{\citenamefont {Danieli}\ \emph {et~al.}(2007)\citenamefont
  {Danieli}, \citenamefont {Álvarez}, \citenamefont {Levstein},\ and\
  \citenamefont {Pastawski}}]{bib:EPQD2}%
  \BibitemOpen
  \bibfield  {author} {\bibinfo {author} {\bibfnamefont {E.}~\bibnamefont
  {Danieli}}, \bibinfo {author} {\bibfnamefont {G.}~\bibnamefont {Álvarez}},
  \bibinfo {author} {\bibfnamefont {P.}~\bibnamefont {Levstein}}, \ and\
  \bibinfo {author} {\bibfnamefont {H.}~\bibnamefont {Pastawski}},\ }\href
  {\doibase http://dx.doi.org/10.1016/j.ssc.2006.11.001} {\bibfield  {journal}
  {\bibinfo  {journal} {Solid State Communications}\ }\textbf {\bibinfo
  {volume} {141}},\ \bibinfo {pages} {422 } (\bibinfo {year}
  {2007})}\BibitemShut {NoStop}%
\bibitem [{\citenamefont {Bender}\ and\ \citenamefont
  {Boettcher}(1998)}]{bib:PTBend98}%
  \BibitemOpen
  \bibfield  {author} {\bibinfo {author} {\bibfnamefont {C.~M.}\ \bibnamefont
  {Bender}}\ and\ \bibinfo {author} {\bibfnamefont {S.}~\bibnamefont
  {Boettcher}},\ }\href {\doibase 10.1103/PhysRevLett.80.5243} {\bibfield
  {journal} {\bibinfo  {journal} {Phys. Rev. Lett.}\ }\textbf {\bibinfo
  {volume} {80}},\ \bibinfo {pages} {5243} (\bibinfo {year}
  {1998})}\BibitemShut {NoStop}%
\bibitem [{\citenamefont {Bender}\ \emph {et~al.}(1999)\citenamefont {Bender},
  \citenamefont {Boettcher},\ and\ \citenamefont {Meisinger}}]{bib:PTBend99}%
  \BibitemOpen
  \bibfield  {author} {\bibinfo {author} {\bibfnamefont {C.~M.}\ \bibnamefont
  {Bender}}, \bibinfo {author} {\bibfnamefont {S.}~\bibnamefont {Boettcher}}, \
  and\ \bibinfo {author} {\bibfnamefont {P.~N.}\ \bibnamefont {Meisinger}},\
  }\href {\doibase http://dx.doi.org/10.1063/1.532860} {\bibfield  {journal}
  {\bibinfo  {journal} {Journal of Mathematical Physics}\ }\textbf {\bibinfo
  {volume} {40}},\ \bibinfo {pages} {2201} (\bibinfo {year}
  {1999})}\BibitemShut {NoStop}%
\bibitem [{\citenamefont {Bender}(2007)}]{bib:PTBend07}%
  \BibitemOpen
  \bibfield  {author} {\bibinfo {author} {\bibfnamefont {C.~M.}\ \bibnamefont
  {Bender}},\ }\href {http://stacks.iop.org/0034-4885/70/i=6/a=R03} {\bibfield
  {journal} {\bibinfo  {journal} {Reports on Progress in Physics}\ }\textbf
  {\bibinfo {volume} {70}},\ \bibinfo {pages} {947} (\bibinfo {year}
  {2007})}\BibitemShut {NoStop}%
\bibitem [{\citenamefont {Guo}\ \emph {et~al.}(2009)\citenamefont {Guo},
  \citenamefont {Salamo}, \citenamefont {Duchesne}, \citenamefont {Morandotti},
  \citenamefont {Volatier-Ravat}, \citenamefont {Aimez}, \citenamefont
  {Siviloglou},\ and\ \citenamefont {Christodoulides}}]{bib:PTopt1}%
  \BibitemOpen
  \bibfield  {author} {\bibinfo {author} {\bibfnamefont {A.}~\bibnamefont
  {Guo}}, \bibinfo {author} {\bibfnamefont {G.~J.}\ \bibnamefont {Salamo}},
  \bibinfo {author} {\bibfnamefont {D.}~\bibnamefont {Duchesne}}, \bibinfo
  {author} {\bibfnamefont {R.}~\bibnamefont {Morandotti}}, \bibinfo {author}
  {\bibfnamefont {M.}~\bibnamefont {Volatier-Ravat}}, \bibinfo {author}
  {\bibfnamefont {V.}~\bibnamefont {Aimez}}, \bibinfo {author} {\bibfnamefont
  {G.~A.}\ \bibnamefont {Siviloglou}}, \ and\ \bibinfo {author} {\bibfnamefont
  {D.~N.}\ \bibnamefont {Christodoulides}},\ }\href {\doibase
  10.1103/PhysRevLett.103.093902} {\bibfield  {journal} {\bibinfo  {journal}
  {Phys. Rev. Lett.}\ }\textbf {\bibinfo {volume} {103}},\ \bibinfo {pages}
  {093902} (\bibinfo {year} {2009})}\BibitemShut {NoStop}%
\bibitem [{\citenamefont {Ruter}\ \emph {et~al.}(2010)\citenamefont {Ruter},
  \citenamefont {Makris}, \citenamefont {El-Ganainy}, \citenamefont
  {Christodoulides}, \citenamefont {Segev},\ and\ \citenamefont
  {Kip}}]{bib:PTOptNat}%
  \BibitemOpen
  \bibfield  {author} {\bibinfo {author} {\bibfnamefont {C.~E.}\ \bibnamefont
  {Ruter}}, \bibinfo {author} {\bibfnamefont {K.~G.}\ \bibnamefont {Makris}},
  \bibinfo {author} {\bibfnamefont {R.}~\bibnamefont {El-Ganainy}}, \bibinfo
  {author} {\bibfnamefont {D.~N.}\ \bibnamefont {Christodoulides}}, \bibinfo
  {author} {\bibfnamefont {M.}~\bibnamefont {Segev}}, \ and\ \bibinfo {author}
  {\bibfnamefont {D.}~\bibnamefont {Kip}},\ }\href {\doibase 10.1038/nphys1515}
  {\bibfield  {journal} {\bibinfo  {journal} {Nat Phys}\ }\textbf {\bibinfo
  {volume} {6}},\ \bibinfo {pages} {192} (\bibinfo {year} {2010})}\BibitemShut
  {NoStop}%
\bibitem [{\citenamefont {Cannata}\ \emph {et~al.}(2007)\citenamefont
  {Cannata}, \citenamefont {Dedonder},\ and\ \citenamefont
  {Ventura}}]{bib:AnnPTScat}%
  \BibitemOpen
  \bibfield  {author} {\bibinfo {author} {\bibfnamefont {F.}~\bibnamefont
  {Cannata}}, \bibinfo {author} {\bibfnamefont {J.-P.}\ \bibnamefont
  {Dedonder}}, \ and\ \bibinfo {author} {\bibfnamefont {A.}~\bibnamefont
  {Ventura}},\ }\href {\doibase http://dx.doi.org/10.1016/j.aop.2006.05.011}
  {\bibfield  {journal} {\bibinfo  {journal} {Annals of Physics}\ }\textbf
  {\bibinfo {volume} {322}},\ \bibinfo {pages} {397} (\bibinfo {year}
  {2007})}\BibitemShut {NoStop}%
\bibitem [{\citenamefont {Hernandez-Coronado}\ \emph
  {et~al.}(2011)\citenamefont {Hernandez-Coronado}, \citenamefont
  {Krej\v{c}i\v{r}\'{i}k},\ and\ \citenamefont {Siegl}}]{bib:PTScat}%
  \BibitemOpen
  \bibfield  {author} {\bibinfo {author} {\bibfnamefont {H.}~\bibnamefont
  {Hernandez-Coronado}}, \bibinfo {author} {\bibfnamefont {D.}~\bibnamefont
  {Krej\v{c}i\v{r}\'{i}k}}, \ and\ \bibinfo {author} {\bibfnamefont
  {P.}~\bibnamefont {Siegl}},\ }\href {\doibase
  http://dx.doi.org/10.1016/j.physleta.2011.04.021} {\bibfield  {journal}
  {\bibinfo  {journal} {Physics Letters A}\ }\textbf {\bibinfo {volume}
  {375}},\ \bibinfo {pages} {2149 } (\bibinfo {year} {2011})}\BibitemShut
  {NoStop}%
\bibitem [{\citenamefont {Jin}\ and\ \citenamefont {Song}(2010)}]{bib:Jin2010}%
  \BibitemOpen
  \bibfield  {author} {\bibinfo {author} {\bibfnamefont {L.}~\bibnamefont
  {Jin}}\ and\ \bibinfo {author} {\bibfnamefont {Z.}~\bibnamefont {Song}},\
  }\href {\doibase 10.1103/PhysRevA.81.032109} {\bibfield  {journal} {\bibinfo
  {journal} {Phys. Rev. A}\ }\textbf {\bibinfo {volume} {81}},\ \bibinfo
  {pages} {032109} (\bibinfo {year} {2010})}\BibitemShut {NoStop}%
\bibitem [{\citenamefont {Jin}\ and\ \citenamefont {Song}(2011)}]{bib:Jin2011}%
  \BibitemOpen
  \bibfield  {author} {\bibinfo {author} {\bibfnamefont {L.}~\bibnamefont
  {Jin}}\ and\ \bibinfo {author} {\bibfnamefont {Z.}~\bibnamefont {Song}},\
  }\href {http://stacks.iop.org/1751-8121/44/i=37/a=375304} {\bibfield
  {journal} {\bibinfo  {journal} {Journal of Physics A: Mathematical and
  Theoretical}\ }\textbf {\bibinfo {volume} {44}},\ \bibinfo {pages} {375304}
  (\bibinfo {year} {2011})}\BibitemShut {NoStop}%
\bibitem [{\citenamefont {Gorbatsevich}\ and\ \citenamefont
  {Shubin}(2016)}]{bib:GorShJETP}%
  \BibitemOpen
  \bibfield  {author} {\bibinfo {author} {\bibfnamefont {A.~A.}\ \bibnamefont
  {Gorbatsevich}}\ and\ \bibinfo {author} {\bibfnamefont {N.~M.}\ \bibnamefont
  {Shubin}},\ }\href {\doibase 10.1134/S0021364016120031} {\bibfield  {journal}
  {\bibinfo  {journal} {JETP Letters}\ }\textbf {\bibinfo {volume} {103}},\
  \bibinfo {pages} {769} (\bibinfo {year} {2016})}\BibitemShut {NoStop}%
\bibitem [{\citenamefont {Kato}(1995)}]{bib:BookKato}%
  \BibitemOpen
  \bibfield  {author} {\bibinfo {author} {\bibfnamefont {T.}~\bibnamefont
  {Kato}},\ }\href {https://books.google.ru/books?id=8ji2kN\_D3BwC} {\emph
  {\bibinfo {title} {Perturbation Theory for Linear Operators}}},\ Classics in
  Mathematics\ (\bibinfo  {publisher} {Springer Berlin Heidelberg},\ \bibinfo
  {year} {1995})\BibitemShut {NoStop}%
\bibitem [{\citenamefont {Heiss}(2004)}]{bib:Heiss2004}%
  \BibitemOpen
  \bibfield  {author} {\bibinfo {author} {\bibfnamefont {W.~D.}\ \bibnamefont
  {Heiss}},\ }\href {http://stacks.iop.org/0305-4470/37/i=6/a=034} {\bibfield
  {journal} {\bibinfo  {journal} {Journal of Physics A: Mathematical and
  General}\ }\textbf {\bibinfo {volume} {37}},\ \bibinfo {pages} {2455}
  (\bibinfo {year} {2004})}\BibitemShut {NoStop}%
\bibitem [{\citenamefont {Heiss}(2012)}]{bib:Heiss2012}%
  \BibitemOpen
  \bibfield  {author} {\bibinfo {author} {\bibfnamefont {W.~D.}\ \bibnamefont
  {Heiss}},\ }\href {http://stacks.iop.org/1751-8121/45/i=44/a=444016}
  {\bibfield  {journal} {\bibinfo  {journal} {Journal of Physics A:
  Mathematical and Theoretical}\ }\textbf {\bibinfo {volume} {45}},\ \bibinfo
  {pages} {444016} (\bibinfo {year} {2012})}\BibitemShut {NoStop}%
\bibitem [{\citenamefont {Berry}(2004)}]{bib:Berry}%
  \BibitemOpen
  \bibfield  {author} {\bibinfo {author} {\bibfnamefont {M.}~\bibnamefont
  {Berry}},\ }\href@noop {} {\bibfield  {journal} {\bibinfo  {journal}
  {Czechoslovak journal of physics}\ }\textbf {\bibinfo {volume} {54}},\
  \bibinfo {pages} {1039} (\bibinfo {year} {2004})}\BibitemShut {NoStop}%
\bibitem [{\citenamefont {Rotter}(2009)}]{bib:Rotter2009}%
  \BibitemOpen
  \bibfield  {author} {\bibinfo {author} {\bibfnamefont {I.}~\bibnamefont
  {Rotter}},\ }\href {http://stacks.iop.org/1751-8121/42/i=15/a=153001}
  {\bibfield  {journal} {\bibinfo  {journal} {Journal of Physics A:
  Mathematical and Theoretical}\ }\textbf {\bibinfo {volume} {42}},\ \bibinfo
  {pages} {153001} (\bibinfo {year} {2009})}\BibitemShut {NoStop}%
\bibitem [{\citenamefont {Rotter}\ and\ \citenamefont
  {Bird}(2015)}]{bib:Rotter2015}%
  \BibitemOpen
  \bibfield  {author} {\bibinfo {author} {\bibfnamefont {I.}~\bibnamefont
  {Rotter}}\ and\ \bibinfo {author} {\bibfnamefont {J.~P.}\ \bibnamefont
  {Bird}},\ }\href {http://stacks.iop.org/0034-4885/78/i=11/a=114001}
  {\bibfield  {journal} {\bibinfo  {journal} {Reports on Progress in Physics}\
  }\textbf {\bibinfo {volume} {78}},\ \bibinfo {pages} {114001} (\bibinfo
  {year} {2015})}\BibitemShut {NoStop}%
\bibitem [{\citenamefont {Wiersig}(2016)}]{bib:EPWier}%
  \BibitemOpen
  \bibfield  {author} {\bibinfo {author} {\bibfnamefont {J.}~\bibnamefont
  {Wiersig}},\ }\href {\doibase 10.1103/PhysRevA.93.033809} {\bibfield
  {journal} {\bibinfo  {journal} {Phys. Rev. A}\ }\textbf {\bibinfo {volume}
  {93}},\ \bibinfo {pages} {033809} (\bibinfo {year} {2016})}\BibitemShut
  {NoStop}%
\bibitem [{\citenamefont {Mostafazadeh}(2002)}]{bib:Mostafa1}%
  \BibitemOpen
  \bibfield  {author} {\bibinfo {author} {\bibfnamefont {A.}~\bibnamefont
  {Mostafazadeh}},\ }\href {\doibase 10.1063/1.1418246} {\bibfield  {journal}
  {\bibinfo  {journal} {Journal of Mathematical Physics}\ }\textbf {\bibinfo
  {volume} {43}},\ \bibinfo {pages} {205} (\bibinfo {year} {2002})},\ \Eprint
  {http://arxiv.org/abs/http://dx.doi.org/10.1063/1.1418246}
  {http://dx.doi.org/10.1063/1.1418246} \BibitemShut {NoStop}%
\bibitem [{\citenamefont {Suchkov}\ \emph {et~al.}(2016)\citenamefont
  {Suchkov}, \citenamefont {Fotsa-Ngaffo}, \citenamefont {Kenfack-Jiotsa},
  \citenamefont {Tikeng}, \citenamefont {Kofane}, \citenamefont {Kivshar},\
  and\ \citenamefont {Sukhorukov}}]{bib:KivshNJP}%
  \BibitemOpen
  \bibfield  {author} {\bibinfo {author} {\bibfnamefont {S.~V.}\ \bibnamefont
  {Suchkov}}, \bibinfo {author} {\bibfnamefont {F.}~\bibnamefont
  {Fotsa-Ngaffo}}, \bibinfo {author} {\bibfnamefont {A.}~\bibnamefont
  {Kenfack-Jiotsa}}, \bibinfo {author} {\bibfnamefont {A.~D.}\ \bibnamefont
  {Tikeng}}, \bibinfo {author} {\bibfnamefont {T.~C.}\ \bibnamefont {Kofane}},
  \bibinfo {author} {\bibfnamefont {Y.~S.}\ \bibnamefont {Kivshar}}, \ and\
  \bibinfo {author} {\bibfnamefont {A.~A.}\ \bibnamefont {Sukhorukov}},\ }\href
  {http://stacks.iop.org/1367-2630/18/i=6/a=065005} {\bibfield  {journal}
  {\bibinfo  {journal} {New Journal of Physics}\ }\textbf {\bibinfo {volume}
  {18}},\ \bibinfo {pages} {065005} (\bibinfo {year} {2016})}\BibitemShut
  {NoStop}%
\bibitem [{\citenamefont {Regensburger}\ \emph {et~al.}(2012)\citenamefont
  {Regensburger}, \citenamefont {Bersch}, \citenamefont {Miri}, \citenamefont
  {Onishchukov}, \citenamefont {Christodoulides},\ and\ \citenamefont
  {Peschel}}]{bib:NatPTLight}%
  \BibitemOpen
  \bibfield  {author} {\bibinfo {author} {\bibfnamefont {A.}~\bibnamefont
  {Regensburger}}, \bibinfo {author} {\bibfnamefont {C.}~\bibnamefont
  {Bersch}}, \bibinfo {author} {\bibfnamefont {M.-A.}\ \bibnamefont {Miri}},
  \bibinfo {author} {\bibfnamefont {G.}~\bibnamefont {Onishchukov}}, \bibinfo
  {author} {\bibfnamefont {D.~N.}\ \bibnamefont {Christodoulides}}, \ and\
  \bibinfo {author} {\bibfnamefont {U.}~\bibnamefont {Peschel}},\ }\href
  {\doibase 10.1038/nature11298} {\bibfield  {journal} {\bibinfo  {journal}
  {Nature}\ }\textbf {\bibinfo {volume} {488}},\ \bibinfo {pages} {167}
  (\bibinfo {year} {2012})}\BibitemShut {NoStop}%
\bibitem [{\citenamefont {Alaeian}\ and\ \citenamefont
  {Dionne}(2014)}]{bib:Alaeian}%
  \BibitemOpen
  \bibfield  {author} {\bibinfo {author} {\bibfnamefont {H.}~\bibnamefont
  {Alaeian}}\ and\ \bibinfo {author} {\bibfnamefont {J.~A.}\ \bibnamefont
  {Dionne}},\ }\href {\doibase 10.1103/PhysRevA.89.033829} {\bibfield
  {journal} {\bibinfo  {journal} {Phys. Rev. A}\ }\textbf {\bibinfo {volume}
  {89}},\ \bibinfo {pages} {033829} (\bibinfo {year} {2014})}\BibitemShut
  {NoStop}%
\bibitem [{\citenamefont {Mostafazadeh}(2016)}]{bib:AnnMeta}%
  \BibitemOpen
  \bibfield  {author} {\bibinfo {author} {\bibfnamefont {A.}~\bibnamefont
  {Mostafazadeh}},\ }\href {\doibase
  http://dx.doi.org/10.1016/j.aop.2016.01.025} {\bibfield  {journal} {\bibinfo
  {journal} {Annals of Physics}\ }\textbf {\bibinfo {volume} {368}},\ \bibinfo
  {pages} {56} (\bibinfo {year} {2016})}\BibitemShut {NoStop}%
\bibitem [{\citenamefont {Liertzer}\ \emph {et~al.}(2012)\citenamefont
  {Liertzer}, \citenamefont {Ge}, \citenamefont {Cerjan}, \citenamefont
  {Stone}, \citenamefont {T\"ureci},\ and\ \citenamefont {Rotter}}]{bib:Laser}%
  \BibitemOpen
  \bibfield  {author} {\bibinfo {author} {\bibfnamefont {M.}~\bibnamefont
  {Liertzer}}, \bibinfo {author} {\bibfnamefont {L.}~\bibnamefont {Ge}},
  \bibinfo {author} {\bibfnamefont {A.}~\bibnamefont {Cerjan}}, \bibinfo
  {author} {\bibfnamefont {A.~D.}\ \bibnamefont {Stone}}, \bibinfo {author}
  {\bibfnamefont {H.~E.}\ \bibnamefont {T\"ureci}}, \ and\ \bibinfo {author}
  {\bibfnamefont {S.}~\bibnamefont {Rotter}},\ }\href {\doibase
  10.1103/PhysRevLett.108.173901} {\bibfield  {journal} {\bibinfo  {journal}
  {Phys. Rev. Lett.}\ }\textbf {\bibinfo {volume} {108}},\ \bibinfo {pages}
  {173901} (\bibinfo {year} {2012})}\BibitemShut {NoStop}%
\bibitem [{\citenamefont {Brandstetter}\ \emph {et~al.}(2014)\citenamefont
  {Brandstetter}, \citenamefont {Liertzer}, \citenamefont {Deutsch},
  \citenamefont {Klang}, \citenamefont {Sch{\"o}berl}, \citenamefont
  {T{\"u}reci}, \citenamefont {Strasser}, \citenamefont {Unterrainer},\ and\
  \citenamefont {Rotter}}]{bib:nature}%
  \BibitemOpen
  \bibfield  {author} {\bibinfo {author} {\bibfnamefont {M.}~\bibnamefont
  {Brandstetter}}, \bibinfo {author} {\bibfnamefont {M.}~\bibnamefont
  {Liertzer}}, \bibinfo {author} {\bibfnamefont {C.}~\bibnamefont {Deutsch}},
  \bibinfo {author} {\bibfnamefont {P.}~\bibnamefont {Klang}}, \bibinfo
  {author} {\bibfnamefont {J.}~\bibnamefont {Sch{\"o}berl}}, \bibinfo {author}
  {\bibfnamefont {H.~E.}\ \bibnamefont {T{\"u}reci}}, \bibinfo {author}
  {\bibfnamefont {G.}~\bibnamefont {Strasser}}, \bibinfo {author}
  {\bibfnamefont {K.}~\bibnamefont {Unterrainer}}, \ and\ \bibinfo {author}
  {\bibfnamefont {S.}~\bibnamefont {Rotter}},\ }\href
  {http://dx.doi.org/10.1038/ncomms5034} {\bibfield  {journal} {\bibinfo
  {journal} {Nat Commun}\ }\textbf {\bibinfo {volume} {5}},\ \bibinfo {pages}
  {4034} (\bibinfo {year} {2014})}\BibitemShut {NoStop}%
\bibitem [{\citenamefont {Longhi}(2010)}]{bib:LonghiCPA}%
  \BibitemOpen
  \bibfield  {author} {\bibinfo {author} {\bibfnamefont {S.}~\bibnamefont
  {Longhi}},\ }\href {\doibase 10.1103/PhysRevA.82.031801} {\bibfield
  {journal} {\bibinfo  {journal} {Phys. Rev. A}\ }\textbf {\bibinfo {volume}
  {82}},\ \bibinfo {pages} {031801} (\bibinfo {year} {2010})}\BibitemShut
  {NoStop}%
\bibitem [{\citenamefont {Heiss}\ and\ \citenamefont
  {Wunner}(2014)}]{bib:Heiss2014}%
  \BibitemOpen
  \bibfield  {author} {\bibinfo {author} {\bibfnamefont {W.~D.}\ \bibnamefont
  {Heiss}}\ and\ \bibinfo {author} {\bibfnamefont {G.}~\bibnamefont {Wunner}},\
  }\href {\doibase 10.1140/epjd/e2014-50379-8} {\bibfield  {journal} {\bibinfo
  {journal} {The European Physical Journal D}\ }\textbf {\bibinfo {volume}
  {68}},\ \bibinfo {pages} {284} (\bibinfo {year} {2014})}\BibitemShut
  {NoStop}%
\bibitem [{\citenamefont {Vanroose}\ \emph {et~al.}(1997)\citenamefont
  {Vanroose}, \citenamefont {Leuven}, \citenamefont {Arickx},\ and\
  \citenamefont {Broeckhove}}]{bib:DoublePole1}%
  \BibitemOpen
  \bibfield  {author} {\bibinfo {author} {\bibfnamefont {W.}~\bibnamefont
  {Vanroose}}, \bibinfo {author} {\bibfnamefont {P.~V.}\ \bibnamefont
  {Leuven}}, \bibinfo {author} {\bibfnamefont {F.}~\bibnamefont {Arickx}}, \
  and\ \bibinfo {author} {\bibfnamefont {J.}~\bibnamefont {Broeckhove}},\
  }\href {http://stacks.iop.org/0305-4470/30/i=15/a=034} {\bibfield  {journal}
  {\bibinfo  {journal} {Journal of Physics A: Mathematical and General}\
  }\textbf {\bibinfo {volume} {30}},\ \bibinfo {pages} {5543} (\bibinfo {year}
  {1997})}\BibitemShut {NoStop}%
\bibitem [{\citenamefont {Ambichl}\ \emph {et~al.}(2013)\citenamefont
  {Ambichl}, \citenamefont {Makris}, \citenamefont {Ge}, \citenamefont {Chong},
  \citenamefont {Stone},\ and\ \citenamefont {Rotter}}]{bib:Ambi}%
  \BibitemOpen
  \bibfield  {author} {\bibinfo {author} {\bibfnamefont {P.}~\bibnamefont
  {Ambichl}}, \bibinfo {author} {\bibfnamefont {K.~G.}\ \bibnamefont {Makris}},
  \bibinfo {author} {\bibfnamefont {L.}~\bibnamefont {Ge}}, \bibinfo {author}
  {\bibfnamefont {Y.}~\bibnamefont {Chong}}, \bibinfo {author} {\bibfnamefont
  {A.~D.}\ \bibnamefont {Stone}}, \ and\ \bibinfo {author} {\bibfnamefont
  {S.}~\bibnamefont {Rotter}},\ }\href {\doibase 10.1103/PhysRevX.3.041030}
  {\bibfield  {journal} {\bibinfo  {journal} {Phys. Rev. X}\ }\textbf {\bibinfo
  {volume} {3}},\ \bibinfo {pages} {041030} (\bibinfo {year}
  {2013})}\BibitemShut {NoStop}%
\bibitem [{\citenamefont {Chong}\ \emph {et~al.}(2011)\citenamefont {Chong},
  \citenamefont {Ge},\ and\ \citenamefont {Stone}}]{bib:Cho}%
  \BibitemOpen
  \bibfield  {author} {\bibinfo {author} {\bibfnamefont {Y.~D.}\ \bibnamefont
  {Chong}}, \bibinfo {author} {\bibfnamefont {L.}~\bibnamefont {Ge}}, \ and\
  \bibinfo {author} {\bibfnamefont {A.~D.}\ \bibnamefont {Stone}},\ }\href
  {\doibase 10.1103/PhysRevLett.106.093902} {\bibfield  {journal} {\bibinfo
  {journal} {Phys. Rev. Lett.}\ }\textbf {\bibinfo {volume} {106}},\ \bibinfo
  {pages} {093902} (\bibinfo {year} {2011})}\BibitemShut {NoStop}%
\bibitem [{\citenamefont {Bulgakov}\ \emph {et~al.}(2006)\citenamefont
  {Bulgakov}, \citenamefont {Pichugin}, \citenamefont {Sadreev},\ and\
  \citenamefont {Rotter}}]{bib:BulgJETP}%
  \BibitemOpen
  \bibfield  {author} {\bibinfo {author} {\bibfnamefont {E.~N.}\ \bibnamefont
  {Bulgakov}}, \bibinfo {author} {\bibfnamefont {K.~N.}\ \bibnamefont
  {Pichugin}}, \bibinfo {author} {\bibfnamefont {A.~F.}\ \bibnamefont
  {Sadreev}}, \ and\ \bibinfo {author} {\bibfnamefont {I.}~\bibnamefont
  {Rotter}},\ }\href {http://dx.doi.org/10.1134/S0021364006200057} {\bibfield
  {journal} {\bibinfo  {journal} {JETP Letters}\ }\textbf {\bibinfo {volume}
  {84}},\ \bibinfo {pages} {430} (\bibinfo {year} {2006})}\BibitemShut
  {NoStop}%
\bibitem [{\citenamefont {Blanchard}\ \emph {et~al.}(2016)\citenamefont
  {Blanchard}, \citenamefont {Hugonin},\ and\ \citenamefont
  {Sauvan}}]{bib:BICDiverg}%
  \BibitemOpen
  \bibfield  {author} {\bibinfo {author} {\bibfnamefont {C.}~\bibnamefont
  {Blanchard}}, \bibinfo {author} {\bibfnamefont {J.-P.}\ \bibnamefont
  {Hugonin}}, \ and\ \bibinfo {author} {\bibfnamefont {C.}~\bibnamefont
  {Sauvan}},\ }\href {\doibase 10.1103/PhysRevB.94.155303} {\bibfield
  {journal} {\bibinfo  {journal} {Phys. Rev. B}\ }\textbf {\bibinfo {volume}
  {94}},\ \bibinfo {pages} {155303} (\bibinfo {year} {2016})}\BibitemShut
  {NoStop}%
\bibitem [{\citenamefont {Okamoto}(2010)}]{bib:okamoto}%
  \BibitemOpen
  \bibfield  {author} {\bibinfo {author} {\bibfnamefont {K.}~\bibnamefont
  {Okamoto}},\ }\href@noop {} {\emph {\bibinfo {title} {Fundamentals of optical
  waveguides}}}\ (\bibinfo  {publisher} {Academic press},\ \bibinfo {year}
  {2010})\BibitemShut {NoStop}%
\bibitem [{\citenamefont {Longhi}(2009{\natexlab{a}})}]{bib:LonghiRev}%
  \BibitemOpen
  \bibfield  {author} {\bibinfo {author} {\bibfnamefont {S.}~\bibnamefont
  {Longhi}},\ }\href {\doibase 10.1002/lpor.200810055} {\bibfield  {journal}
  {\bibinfo  {journal} {Laser \& Photonics Reviews}\ }\textbf {\bibinfo
  {volume} {3}},\ \bibinfo {pages} {243} (\bibinfo {year}
  {2009}{\natexlab{a}})}\BibitemShut {NoStop}%
\bibitem [{\citenamefont {Caroli}\ \emph {et~al.}(1971)\citenamefont {Caroli},
  \citenamefont {Combescot}, \citenamefont {Nozieres},\ and\ \citenamefont
  {Saint-James}}]{bib:Caroli1971}%
  \BibitemOpen
  \bibfield  {author} {\bibinfo {author} {\bibfnamefont {C.}~\bibnamefont
  {Caroli}}, \bibinfo {author} {\bibfnamefont {R.}~\bibnamefont {Combescot}},
  \bibinfo {author} {\bibfnamefont {P.}~\bibnamefont {Nozieres}}, \ and\
  \bibinfo {author} {\bibfnamefont {D.}~\bibnamefont {Saint-James}},\ }\href
  {http://stacks.iop.org/0022-3719/4/i=8/a=018} {\bibfield  {journal} {\bibinfo
   {journal} {Journal of Physics C: Solid State Physics}\ }\textbf {\bibinfo
  {volume} {4}},\ \bibinfo {pages} {916} (\bibinfo {year} {1971})}\BibitemShut
  {NoStop}%
\bibitem [{\citenamefont {Datta}(1997)}]{bib:DattaBook1997}%
  \BibitemOpen
  \bibfield  {author} {\bibinfo {author} {\bibfnamefont {S.}~\bibnamefont
  {Datta}},\ }\href {https://books.google.fr/books?id=28BC-ofEhvUC} {\emph
  {\bibinfo {title} {Electronic Transport in Mesoscopic Systems}}},\ Cambridge
  Studies in Semiconductor Physi\ (\bibinfo  {publisher} {Cambridge University
  Press},\ \bibinfo {year} {1997})\BibitemShut {NoStop}%
\bibitem [{\citenamefont {Sokolov}\ and\ \citenamefont
  {Zelevinsky}(1992)}]{bib:HeffZelv}%
  \BibitemOpen
  \bibfield  {author} {\bibinfo {author} {\bibfnamefont {V.~V.}\ \bibnamefont
  {Sokolov}}\ and\ \bibinfo {author} {\bibfnamefont {V.~G.}\ \bibnamefont
  {Zelevinsky}},\ }\href@noop {} {\bibfield  {journal} {\bibinfo  {journal}
  {Annals of Physics}\ }\textbf {\bibinfo {volume} {216}},\ \bibinfo {pages}
  {323} (\bibinfo {year} {1992})}\BibitemShut {NoStop}%
\bibitem [{\citenamefont {Sherman}\ and\ \citenamefont
  {Morrison}(1950)}]{bib:SherMorr}%
  \BibitemOpen
  \bibfield  {author} {\bibinfo {author} {\bibfnamefont {J.}~\bibnamefont
  {Sherman}}\ and\ \bibinfo {author} {\bibfnamefont {W.~J.}\ \bibnamefont
  {Morrison}},\ }\href {http://www.jstor.org/stable/2236561} {\bibfield
  {journal} {\bibinfo  {journal} {The Annals of Mathematical Statistics}\
  }\textbf {\bibinfo {volume} {21}},\ \bibinfo {pages} {124} (\bibinfo {year}
  {1950})}\BibitemShut {NoStop}%
\bibitem [{\citenamefont {Harville}(1997)}]{bib:MatDetLem}%
  \BibitemOpen
  \bibfield  {author} {\bibinfo {author} {\bibfnamefont {D.~A.}\ \bibnamefont
  {Harville}},\ }\href@noop {} {\emph {\bibinfo {title} {Matrix algebra from a
  statistician's perspective}}},\ Vol.~\bibinfo {volume} {1}\ (\bibinfo
  {publisher} {Springer},\ \bibinfo {year} {1997})\BibitemShut {NoStop}%
\bibitem [{\citenamefont {Fano}(1961)}]{bib:Fano1961}%
  \BibitemOpen
  \bibfield  {author} {\bibinfo {author} {\bibfnamefont {U.}~\bibnamefont
  {Fano}},\ }\href {\doibase 10.1103/PhysRev.124.1866} {\bibfield  {journal}
  {\bibinfo  {journal} {Phys. Rev.}\ }\textbf {\bibinfo {volume} {124}},\
  \bibinfo {pages} {1866} (\bibinfo {year} {1961})}\BibitemShut {NoStop}%
\bibitem [{\citenamefont {Tanaka}\ \emph {et~al.}(2016)\citenamefont {Tanaka},
  \citenamefont {Garmon}, \citenamefont {Kanki},\ and\ \citenamefont
  {Petrosky}}]{bib:FanoEP}%
  \BibitemOpen
  \bibfield  {author} {\bibinfo {author} {\bibfnamefont {S.}~\bibnamefont
  {Tanaka}}, \bibinfo {author} {\bibfnamefont {S.}~\bibnamefont {Garmon}},
  \bibinfo {author} {\bibfnamefont {K.}~\bibnamefont {Kanki}}, \ and\ \bibinfo
  {author} {\bibfnamefont {T.}~\bibnamefont {Petrosky}},\ }\href {\doibase
  10.1103/PhysRevA.94.022105} {\bibfield  {journal} {\bibinfo  {journal} {Phys.
  Rev. A}\ }\textbf {\bibinfo {volume} {94}},\ \bibinfo {pages} {022105}
  (\bibinfo {year} {2016})}\BibitemShut {NoStop}%
\bibitem [{\citenamefont {Lu}\ \emph {et~al.}(2005)\citenamefont {Lu},
  \citenamefont {L\"u},\ and\ \citenamefont {Zhu}}]{bib:ABLu}%
  \BibitemOpen
  \bibfield  {author} {\bibinfo {author} {\bibfnamefont {H.}~\bibnamefont
  {Lu}}, \bibinfo {author} {\bibfnamefont {R.}~\bibnamefont {L\"u}}, \ and\
  \bibinfo {author} {\bibfnamefont {B.-f.}\ \bibnamefont {Zhu}},\ }\href
  {\doibase 10.1103/PhysRevB.71.235320} {\bibfield  {journal} {\bibinfo
  {journal} {Phys. Rev. B}\ }\textbf {\bibinfo {volume} {71}},\ \bibinfo
  {pages} {235320} (\bibinfo {year} {2005})}\BibitemShut {NoStop}%
\bibitem [{\citenamefont {Ladr\'on~de Guevara}\ and\ \citenamefont
  {Orellana}(2006)}]{bib:ABQD1}%
  \BibitemOpen
  \bibfield  {author} {\bibinfo {author} {\bibfnamefont {M.~L.}\ \bibnamefont
  {Ladr\'on~de Guevara}}\ and\ \bibinfo {author} {\bibfnamefont {P.~A.}\
  \bibnamefont {Orellana}},\ }\href {\doibase 10.1103/PhysRevB.73.205303}
  {\bibfield  {journal} {\bibinfo  {journal} {Phys. Rev. B}\ }\textbf {\bibinfo
  {volume} {73}},\ \bibinfo {pages} {205303} (\bibinfo {year}
  {2006})}\BibitemShut {NoStop}%
\bibitem [{\citenamefont {Gong}\ \emph {et~al.}(2009)\citenamefont {Gong},
  \citenamefont {Han},\ and\ \citenamefont {Wei}}]{bib:ABQD2}%
  \BibitemOpen
  \bibfield  {author} {\bibinfo {author} {\bibfnamefont {W.}~\bibnamefont
  {Gong}}, \bibinfo {author} {\bibfnamefont {Y.}~\bibnamefont {Han}}, \ and\
  \bibinfo {author} {\bibfnamefont {G.}~\bibnamefont {Wei}},\ }\href
  {http://stacks.iop.org/0953-8984/21/i=17/a=175801} {\bibfield  {journal}
  {\bibinfo  {journal} {Journal of Physics: Condensed Matter}\ }\textbf
  {\bibinfo {volume} {21}},\ \bibinfo {pages} {175801} (\bibinfo {year}
  {2009})}\BibitemShut {NoStop}%
\bibitem [{\citenamefont {Yan}\ and\ \citenamefont {Fu}(2013)}]{bib:ABQD3}%
  \BibitemOpen
  \bibfield  {author} {\bibinfo {author} {\bibfnamefont {J.-X.}\ \bibnamefont
  {Yan}}\ and\ \bibinfo {author} {\bibfnamefont {H.-H.}\ \bibnamefont {Fu}},\
  }\href {\doibase https://doi.org/10.1016/j.physb.2012.11.009} {\bibfield
  {journal} {\bibinfo  {journal} {Physica B: Condensed Matter}\ }\textbf
  {\bibinfo {volume} {410}},\ \bibinfo {pages} {197} (\bibinfo {year}
  {2013})}\BibitemShut {NoStop}%
\bibitem [{\citenamefont {Ryndyk}\ \emph {et~al.}(2009)\citenamefont {Ryndyk},
  \citenamefont {Guti{\'e}rrez}, \citenamefont {Song},\ and\ \citenamefont
  {Cuniberti}}]{bib:tightBindSelfE}%
  \BibitemOpen
  \bibfield  {author} {\bibinfo {author} {\bibfnamefont {D.}~\bibnamefont
  {Ryndyk}}, \bibinfo {author} {\bibfnamefont {R.}~\bibnamefont
  {Guti{\'e}rrez}}, \bibinfo {author} {\bibfnamefont {B.}~\bibnamefont {Song}},
  \ and\ \bibinfo {author} {\bibfnamefont {G.}~\bibnamefont {Cuniberti}},\ }in\
  \href@noop {} {\emph {\bibinfo {booktitle} {Energy Transfer Dynamics in
  Biomaterial Systems}}}\ (\bibinfo  {publisher} {Springer},\ \bibinfo {year}
  {2009})\ pp.\ \bibinfo {pages} {213--335}\BibitemShut {NoStop}%
\bibitem [{\citenamefont {Smith}(2011)}]{bib:Aroma2}%
  \BibitemOpen
  \bibfield  {author} {\bibinfo {author} {\bibfnamefont {M.}~\bibnamefont
  {Smith}},\ }\href {https://books.google.ru/books?id=gKPtYs0jyMcC} {\emph
  {\bibinfo {title} {Organic Chemistry: An Acid—Base Approach}}}\ (\bibinfo
  {publisher} {Taylor \& Francis},\ \bibinfo {year} {2011})\BibitemShut
  {NoStop}%
\bibitem [{\citenamefont {L\"{o}wdin}(1950)}]{bib:Lowdin}%
  \BibitemOpen
  \bibfield  {author} {\bibinfo {author} {\bibfnamefont {P.}~\bibnamefont
  {L\"{o}wdin}},\ }\href {\doibase 10.1063/1.1747632} {\bibfield  {journal}
  {\bibinfo  {journal} {The Journal of Chemical Physics}\ }\textbf {\bibinfo
  {volume} {18}},\ \bibinfo {pages} {365} (\bibinfo {year} {1950})},\ \Eprint
  {http://arxiv.org/abs/http://dx.doi.org/10.1063/1.1747632}
  {http://dx.doi.org/10.1063/1.1747632} \BibitemShut {NoStop}%
\bibitem [{\citenamefont {L\"{o}wdin}(1962)}]{bib:Lowdin1962}%
  \BibitemOpen
  \bibfield  {author} {\bibinfo {author} {\bibfnamefont {P.}~\bibnamefont
  {L\"{o}wdin}},\ }\href {\doibase 10.1063/1.1724312} {\bibfield  {journal}
  {\bibinfo  {journal} {Journal of Mathematical Physics}\ }\textbf {\bibinfo
  {volume} {3}},\ \bibinfo {pages} {969} (\bibinfo {year} {1962})},\ \Eprint
  {http://arxiv.org/abs/http://dx.doi.org/10.1063/1.1724312}
  {http://dx.doi.org/10.1063/1.1724312} \BibitemShut {NoStop}%
\bibitem [{\citenamefont {L\"{o}wdin}(1963)}]{bib:Lowdin1963}%
  \BibitemOpen
  \bibfield  {author} {\bibinfo {author} {\bibfnamefont {P.-O.}\ \bibnamefont
  {L\"{o}wdin}},\ }\href {\doibase
  http://dx.doi.org/10.1016/0022-2852(63)90151-6} {\bibfield  {journal}
  {\bibinfo  {journal} {Journal of Molecular Spectroscopy}\ }\textbf {\bibinfo
  {volume} {10}},\ \bibinfo {pages} {12} (\bibinfo {year} {1963})}\BibitemShut
  {NoStop}%
\bibitem [{\citenamefont {Vaupel}\ \emph {et~al.}(1996)\citenamefont {Vaupel},
  \citenamefont {Thomas}, \citenamefont {K\"uhn}, \citenamefont {May},
  \citenamefont {Maschke}, \citenamefont {Heberle}, \citenamefont {R\"uhle},\
  and\ \citenamefont {K\"ohler}}]{bib:Thomas}%
  \BibitemOpen
  \bibfield  {author} {\bibinfo {author} {\bibfnamefont {H.}~\bibnamefont
  {Vaupel}}, \bibinfo {author} {\bibfnamefont {P.}~\bibnamefont {Thomas}},
  \bibinfo {author} {\bibfnamefont {O.}~\bibnamefont {K\"uhn}}, \bibinfo
  {author} {\bibfnamefont {V.}~\bibnamefont {May}}, \bibinfo {author}
  {\bibfnamefont {K.}~\bibnamefont {Maschke}}, \bibinfo {author} {\bibfnamefont
  {A.~P.}\ \bibnamefont {Heberle}}, \bibinfo {author} {\bibfnamefont {W.~W.}\
  \bibnamefont {R\"uhle}}, \ and\ \bibinfo {author} {\bibfnamefont
  {K.}~\bibnamefont {K\"ohler}},\ }\href {\doibase 10.1103/PhysRevB.53.16531}
  {\bibfield  {journal} {\bibinfo  {journal} {Phys. Rev. B}\ }\textbf {\bibinfo
  {volume} {53}},\ \bibinfo {pages} {16531} (\bibinfo {year}
  {1996})}\BibitemShut {NoStop}%
\bibitem [{\citenamefont {Longhi}(2009{\natexlab{b}})}]{bib:Longhi2009}%
  \BibitemOpen
  \bibfield  {author} {\bibinfo {author} {\bibfnamefont {S.}~\bibnamefont
  {Longhi}},\ }\href {\doibase 10.1080/09500340802187373} {\bibfield  {journal}
  {\bibinfo  {journal} {Journal of Modern Optics}\ }\textbf {\bibinfo {volume}
  {56}},\ \bibinfo {pages} {729} (\bibinfo {year}
  {2009}{\natexlab{b}})}\BibitemShut {NoStop}%
\bibitem [{\citenamefont {Schiegg}\ \emph {et~al.}(2017)\citenamefont
  {Schiegg}, \citenamefont {Dzierzawa},\ and\ \citenamefont
  {Eckern}}]{bib:RectTrans}%
  \BibitemOpen
  \bibfield  {author} {\bibinfo {author} {\bibfnamefont {C.~H.}\ \bibnamefont
  {Schiegg}}, \bibinfo {author} {\bibfnamefont {M.}~\bibnamefont {Dzierzawa}},
  \ and\ \bibinfo {author} {\bibfnamefont {U.}~\bibnamefont {Eckern}},\ }\href
  {http://stacks.iop.org/0953-8984/29/i=8/a=085303} {\bibfield  {journal}
  {\bibinfo  {journal} {Journal of Physics: Condensed Matter}\ }\textbf
  {\bibinfo {volume} {29}},\ \bibinfo {pages} {085303} (\bibinfo {year}
  {2017})}\BibitemShut {NoStop}%
\bibitem [{bib(2014)}]{bib:QuantHeat}%
  \BibitemOpen
  \href {\doibase 10.1146/annurev-physchem-040513-103724} {\bibfield  {journal}
  {\bibinfo  {journal} {Annual Review of Physical Chemistry}\ }\textbf
  {\bibinfo {volume} {65}},\ \bibinfo {pages} {365} (\bibinfo {year} {2014})},\
  \bibinfo {note} {pMID: 24689798},\ \Eprint
  {http://arxiv.org/abs/http://dx.doi.org/10.1146/annurev-physchem-040513-103724}
  {http://dx.doi.org/10.1146/annurev-physchem-040513-103724} \BibitemShut
  {NoStop}%
\bibitem [{\citenamefont {Whitney}(2015)}]{bib:QHEPRB}%
  \BibitemOpen
  \bibfield  {author} {\bibinfo {author} {\bibfnamefont {R.~S.}\ \bibnamefont
  {Whitney}},\ }\href {\doibase 10.1103/PhysRevB.91.115425} {\bibfield
  {journal} {\bibinfo  {journal} {Phys. Rev. B}\ }\textbf {\bibinfo {volume}
  {91}},\ \bibinfo {pages} {115425} (\bibinfo {year} {2015})}\BibitemShut
  {NoStop}%
\bibitem [{\citenamefont {Menzel}\ \emph {et~al.}(2003)\citenamefont {Menzel},
  \citenamefont {Zhu}, \citenamefont {Wu},\ and\ \citenamefont
  {Bogelsack}}]{bib:Filt2}%
  \BibitemOpen
  \bibfield  {author} {\bibinfo {author} {\bibfnamefont {W.}~\bibnamefont
  {Menzel}}, \bibinfo {author} {\bibfnamefont {L.}~\bibnamefont {Zhu}},
  \bibinfo {author} {\bibfnamefont {K.}~\bibnamefont {Wu}}, \ and\ \bibinfo
  {author} {\bibfnamefont {F.}~\bibnamefont {Bogelsack}},\ }\href {\doibase
  10.1109/TMTT.2002.807843} {\bibfield  {journal} {\bibinfo  {journal} {IEEE
  Transactions on Microwave Theory and Techniques}\ }\textbf {\bibinfo {volume}
  {51}},\ \bibinfo {pages} {364} (\bibinfo {year} {2003})}\BibitemShut
  {NoStop}%
\bibitem [{\citenamefont {Zhou}\ \emph {et~al.}(2015)\citenamefont {Zhou},
  \citenamefont {Yin}, \citenamefont {Zhang}, \citenamefont {Chen},\ and\
  \citenamefont {Li}}]{bib:Filt3}%
  \BibitemOpen
  \bibfield  {author} {\bibinfo {author} {\bibfnamefont {X.}~\bibnamefont
  {Zhou}}, \bibinfo {author} {\bibfnamefont {X.}~\bibnamefont {Yin}}, \bibinfo
  {author} {\bibfnamefont {T.}~\bibnamefont {Zhang}}, \bibinfo {author}
  {\bibfnamefont {L.}~\bibnamefont {Chen}}, \ and\ \bibinfo {author}
  {\bibfnamefont {X.}~\bibnamefont {Li}},\ }\href {\doibase
  10.1364/OE.23.011657} {\bibfield  {journal} {\bibinfo  {journal} {Opt.
  Express}\ }\textbf {\bibinfo {volume} {23}},\ \bibinfo {pages} {11657}
  (\bibinfo {year} {2015})}\BibitemShut {NoStop}%
\bibitem [{\citenamefont {Joannopoulos}\ \emph {et~al.}(1997)\citenamefont
  {Joannopoulos}, \citenamefont {Villeneuve},\ and\ \citenamefont
  {Fan}}]{bib:PhotCrNat}%
  \BibitemOpen
  \bibfield  {author} {\bibinfo {author} {\bibfnamefont {J.~D.}\ \bibnamefont
  {Joannopoulos}}, \bibinfo {author} {\bibfnamefont {P.~R.}\ \bibnamefont
  {Villeneuve}}, \ and\ \bibinfo {author} {\bibfnamefont {S.}~\bibnamefont
  {Fan}},\ }\href {\doibase 10.1038/386143a0} {\bibfield  {journal} {\bibinfo
  {journal} {Nature}\ }\textbf {\bibinfo {volume} {386}},\ \bibinfo {pages}
  {143} (\bibinfo {year} {1997})}\BibitemShut {NoStop}%
\bibitem [{\citenamefont {Yu}\ \emph {et~al.}(2015)\citenamefont {Yu},
  \citenamefont {Chen}, \citenamefont {Hu}, \citenamefont {Xue}, \citenamefont
  {Yvind},\ and\ \citenamefont {Mork}}]{bib:PhotCrFano}%
  \BibitemOpen
  \bibfield  {author} {\bibinfo {author} {\bibfnamefont {Y.}~\bibnamefont
  {Yu}}, \bibinfo {author} {\bibfnamefont {Y.}~\bibnamefont {Chen}}, \bibinfo
  {author} {\bibfnamefont {H.}~\bibnamefont {Hu}}, \bibinfo {author}
  {\bibfnamefont {W.}~\bibnamefont {Xue}}, \bibinfo {author} {\bibfnamefont
  {K.}~\bibnamefont {Yvind}}, \ and\ \bibinfo {author} {\bibfnamefont
  {J.}~\bibnamefont {Mork}},\ }\href {\doibase 10.1002/lpor.201400207}
  {\bibfield  {journal} {\bibinfo  {journal} {Laser \& Photonics Reviews}\
  }\textbf {\bibinfo {volume} {9}},\ \bibinfo {pages} {241} (\bibinfo {year}
  {2015})}\BibitemShut {NoStop}%
\bibitem [{\citenamefont {Mingaleev}\ and\ \citenamefont
  {Kivshar}(2002)}]{bib:Kivsh1}%
  \BibitemOpen
  \bibfield  {author} {\bibinfo {author} {\bibfnamefont {S.~F.}\ \bibnamefont
  {Mingaleev}}\ and\ \bibinfo {author} {\bibfnamefont {Y.~S.}\ \bibnamefont
  {Kivshar}},\ }\href@noop {} {\bibfield  {journal} {\bibinfo  {journal}
  {Optics letters}\ }\textbf {\bibinfo {volume} {27}},\ \bibinfo {pages} {231}
  (\bibinfo {year} {2002})}\BibitemShut {NoStop}%
\bibitem [{\citenamefont {Mingaleev}\ \emph {et~al.}(2000)\citenamefont
  {Mingaleev}, \citenamefont {Kivshar},\ and\ \citenamefont
  {Sammut}}]{bib:Kivsh2}%
  \BibitemOpen
  \bibfield  {author} {\bibinfo {author} {\bibfnamefont {S.~F.}\ \bibnamefont
  {Mingaleev}}, \bibinfo {author} {\bibfnamefont {Y.~S.}\ \bibnamefont
  {Kivshar}}, \ and\ \bibinfo {author} {\bibfnamefont {R.~A.}\ \bibnamefont
  {Sammut}},\ }\href@noop {} {\bibfield  {journal} {\bibinfo  {journal}
  {Physical Review E}\ }\textbf {\bibinfo {volume} {62}},\ \bibinfo {pages}
  {5777} (\bibinfo {year} {2000})}\BibitemShut {NoStop}%
\bibitem [{\citenamefont {Flach}\ \emph {et~al.}(2003)\citenamefont {Flach},
  \citenamefont {Miroshnichenko}, \citenamefont {Fleurov},\ and\ \citenamefont
  {Fistul}}]{bib:Mir1}%
  \BibitemOpen
  \bibfield  {author} {\bibinfo {author} {\bibfnamefont {S.}~\bibnamefont
  {Flach}}, \bibinfo {author} {\bibfnamefont {A.}~\bibnamefont
  {Miroshnichenko}}, \bibinfo {author} {\bibfnamefont {V.}~\bibnamefont
  {Fleurov}}, \ and\ \bibinfo {author} {\bibfnamefont {M.}~\bibnamefont
  {Fistul}},\ }\href@noop {} {\bibfield  {journal} {\bibinfo  {journal}
  {Physical review letters}\ }\textbf {\bibinfo {volume} {90}},\ \bibinfo
  {pages} {084101} (\bibinfo {year} {2003})}\BibitemShut {NoStop}%
\bibitem [{\citenamefont {Miroshnichenko}\ and\ \citenamefont
  {Kivshar}(2005)}]{bib:TransZeros}%
  \BibitemOpen
  \bibfield  {author} {\bibinfo {author} {\bibfnamefont {A.~E.}\ \bibnamefont
  {Miroshnichenko}}\ and\ \bibinfo {author} {\bibfnamefont {Y.~S.}\
  \bibnamefont {Kivshar}},\ }\href {\doibase 10.1103/PhysRevE.72.056611}
  {\bibfield  {journal} {\bibinfo  {journal} {Phys. Rev. E}\ }\textbf {\bibinfo
  {volume} {72}},\ \bibinfo {pages} {056611} (\bibinfo {year}
  {2005})}\BibitemShut {NoStop}%
\bibitem [{\citenamefont {Noble}\ and\ \citenamefont
  {Daniel}(1977)}]{bib:BlMatInv}%
  \BibitemOpen
  \bibfield  {author} {\bibinfo {author} {\bibfnamefont {B.}~\bibnamefont
  {Noble}}\ and\ \bibinfo {author} {\bibfnamefont {J.}~\bibnamefont {Daniel}},\
  }\href {https://books.google.com.au/books?id=jQTvAAAAMAAJ} {\emph {\bibinfo
  {title} {Applied linear algebra}}}\ (\bibinfo  {publisher} {Prentice Hall},\
  \bibinfo {year} {1977})\BibitemShut {NoStop}%
\end{thebibliography}%

\end{document}